\documentclass[5p]{elsarticle}
\makeatletter
\def\ps@pprintTitle{%
	\let\@oddhead\@empty
	\let\@evenhead\@empty
	\def\@oddfoot{\footnotesize\itshape
		\begin{minipage}{9cm}
			\vspace{2mm}
			This is a preprint of an article accepted for publication in the Journal of Systems and Software. Please cite as 'authors. title. Journal of Systems and Software. 2022 (to appear)' 
		\end{minipage}
	}%
	\let\@evenfoot\@oddfoot}

\usepackage{soul}
\newcommand{\revised}[1]{#1}
\newcommand{\rerevised}[1]{#1}

\usepackage{hyperref}
 
\journal{Journal of Systems and Software}

\usepackage{amsmath,amssymb,amsfonts}
\usepackage{graphicx}
\usepackage{xcolor}
\usepackage{booktabs}

\usepackage{amsthm}
\theoremstyle{definition}
\newtheorem{example}{Example}

\usepackage{algorithm} 
\usepackage{algpseudocode} 
\makeatletter
\newcommand*{\algrule}[1][\algorithmicindent]{%
  \makebox[#1][l]{%
    \hspace*{.1em}
    \vrule height .75\baselineskip depth .25\baselineskip
  }
}

\newcount\ALG@printindent@tempcnta
\def\ALG@printindent{%
    \ifnum \theALG@nested>0
    \ifx\ALG@text\ALG@x@notext
    \else
    \unskip
    \ALG@printindent@tempcnta=1
    \loop
    \algrule[\csname ALG@ind@\the\ALG@printindent@tempcnta\endcsname]%
    \advance \ALG@printindent@tempcnta 1
    \ifnum \ALG@printindent@tempcnta<\numexpr\theALG@nested+1\relax
    \repeat
    \fi
    \fi
}
\makeatother
\algrenewcommand\algorithmicindent{1.1em}%

\newcommand{\acronym}{VERACITY}

\usepackage{setspace}

\newtheorem{definition2}{Definition}

\usepackage{framed,caption}
\colorlet{shadecolor}{yellow}

\usepackage{balance}

\begin{document}
\begin{frontmatter}

\title{Quantitative Verification with Adaptive Uncertainty Reduction}

\author{Naif Alasmari$^1$, Radu Calinescu$^1$, Colin Paterson$^1$, and Raffaela Mirandola$^2$}

\address{1. Department of Computer Science, University of York, UK \\
2. Politecnico di Milano, Italy}

\begin{singlespace}
\begin{abstract}
Stochastic models are widely used to verify whether systems satisfy their reliability, performance and other nonfunctional requirements. However, the validity of the verification depends on how accurately the parameters of these models can be estimated using data from component unit testing, monitoring, system logs, etc. When insufficient data are available, the models are affected by epistemic parametric uncertainty, the verification results are inaccurate, and any engineering decisions based on them may be invalid. To address these problems, 
we introduce \acronym, a tool-supported iterative approach for the efficient and accurate verification of nonfunctional requirements under epistemic parameter uncertainty. \acronym\ integrates confidence-interval quantitative verification with a new \emph{adaptive uncertainty reduction} heuristic that collects additional data about the parameters of the verified model by unit-testing specific system components over a series of verification iterations. \acronym\ supports the quantitative verification of \revised{discrete-time} Markov chains, 
deciding which components are to be tested in each iteration based on factors that include the sensitivity of the model to variations in the parameters of different components, and the  overheads (e.g., time or cost) of unit-testing each of these components. We show the effectiveness and efficiency of \acronym\ by using it for the verification of the nonfunctional requirements of a tele-assistance service-based system and an online shopping web application.
\end{abstract}

\end{singlespace}
\end{frontmatter}

\section{Introduction}

\rerevised{Many software systems must satisfy} dependability, performance, cost and other requirements not directly related to \rerevised{the functionality they provide. 
These requirements are termed} \emph{nonfunctional requirements}~\cite{8665968,chung2012non}, and \rerevised{their verification} needs to consider the stochastic nature of software characteristics such as inputs, workloads, timeouts and failures. As such, stochastic \rerevised{models} ranging from  Markov chains \cite{calinescu2011dynamic,gallotti2008quality} and probabilistic automata \cite{Krka:2010:PAA:1808877.1808881,johnson2013incremental} to stochastic Petri nets \cite{marsan-etal1994,perez-etal2012} are widely used to perform this verification. However, ensuring that stochastic models are sufficiently accurate \rerevised{to support this} verification is very challenging. While the structure of the models can be extracted from the actual code \cite{DBLP:conf/icse/FilieriPV13} or from software artefacts such as activity diagrams \cite{gallotti2008quality,6693145}, \rerevised{their} parameters \rerevised{(e.g.,} probabilities, timing and other quantitative information\rerevised{)} are affected by uncertainty. 

These parameters need to be estimated using data obtained, for instance, from testing the system components individually, or (for systems already in use) from system logs. Point estimators such as the mean of the observed parameter values are typically used for this purpose. 
However, the point estimation of the uncertain model parameters produces imprecise verification results, and risks causing invalid engineering decisions~\cite{calinescu2015formal,8428471}, especially when only few observations are available. 
In mature subjects like medicine \cite{akobeng2008confidence,kirkwood2010essential} and in established engineering disciplines like civil \cite{benjamin2014probability} and mechanical \cite{aughenbaugh2006value} engineering, this risk is deemed unacceptable, and \rerevised{it} is mitigated by computing 
\emph{confidence intervals} for the model parameters and the verified properties~\cite{du2009confidence,Gardner746}. 
In contrast, this risk is rarely considered in software performance and dependability engineering. Instead, the research in this area focuses on devising new techniques, tools and applications for the verification of stochastic models, under the strong assumption that using point estimates for the model parameters is sufficiently accurate.  

To address the risk of invalid decisions associated with this assumption, we introduce  \acronym,\footnote{\scalebox{0.97}[1]{quantitative VERification with Adaptive unCertaInTY reduction}} a tool-sup\-port\-ed approach for the quantitative verification of \revised{discrete-time} Markov chains under epistemic parametric \revised{(i.e., transition probability)} uncertainty. Uncertainty is termed \emph{epistemic} when it is due to insufficient data (and therefore reducible by gathering additional data), and \emph{aleatory} when it is intrinsic to the analysed system (and therefore irreducible)~\cite{ICPE14}.

\acronym\ builds on our previous research on computing confidence intervals for the reliability, performance and other nonfunctional properties of a system~\cite{calinescu2015formal,calinescu2016fact}. This computation uses a \emph{parametric \revised{discrete-time} Markov chain} (i.e., a \revised{discrete-time} Markov chain with unknown state transition probabilities) that models the system behaviour, and observations of the system behaviour available from component unit testing, runtime monitoring or system logs. However, when insufficient observations are available, these confidence intervals are too wide to verify whether nonfunctional requirements that impose constraints on such properties are satisfied. 
As an example, given too few observations, the $99\%$ confidence interval for the probability that user requests are handled successfully by a web server may be $[0.76,0.98]$,\footnote{In the case when no observations are available (corresponding to the common scenario of a system that has not yet been tested), this confidence interval is $[0,1]$.} which is too wide for verifying the requirement \emph{`The web server shall handle user requests with a success probability of at least 0.92 at 99\% confidence level.'} To handle this frequently encountered problem efficiently, \acronym\ obtains additional observations by unit-testing specific system components over a series of \emph{adaptive uncertainty reduction} iterations. The components tested in each iteration are decided using a heuristic that takes into account multiple factors. These factors are detailed later in the paper, and include the sensitivity of the nonfunctional properties to variations in the parameters associated with different components, and the overheads (e.g., time or cost) of testing each of these components.

\acronym\ supports both the verification of new system designs, and the verification of planned updates to existing systems. Using \acronym\ to decide whether a new system should be deployed or not involves applying our approach with few or no initial observations of the system parameters. Multiple uncertainty reduction iterations are typically required to acquire sufficient observations of the parameters of the system and to reach a decision in this case. In contrast, when deciding whether updated versions of specific system components should be adopted, \acronym\ can exploit a large number of initial observations of the parameters associated with the components not being updated. Accordingly, fewer uncertainty reduction iterations are typically necessary in this case, primarily to acquire observations of the parameters associated with the updated components.

\revised{VERACITY relies on the possibility to test the components of a system individually. Such unit testing of software components is widely used in software development~}\cite{6982627}\revised{. For certain types of software systems, the individual testing of their components is also feasible after deployment. As an example, the third-party services used by a service-based system can be invoked independently at any stage of the software development life cycle.}

The contributions of the paper are threefold. First, we introduce a new heuristic for the efficient reduction of epistemic parametric uncertainty of Markov chains used in dependability and performance software engineering. Second, we present a new approach that integrates this heuristic with a recently proposed method for formal verification with confidence intervals~\cite{calinescu2015formal,calinescu2016fact}, and a tool that implements the approach, automating the verification of nonfunctional requirements under parametric uncertainty. Finally, we present extensive experimental results showing: (i)~the effectiveness of the  \acronym\ verification approach during initial software development and software updating; and (ii)~the efficiency of the \acronym\ uncertainty reduction compared to uncertainty reduction by uniformly testing all the components of the system under verification (SUV).

We organised the remainder of the paper as follows. Section~\ref{sect:preliminaries} defines the probabilistic model checking and formal verification with confidence concepts and notation required to present \acronym. Section~\ref{sect:example} introduces a motivating example that we then use to present our quantitative verification approach in Section~\ref{sect:approach}. Sections~\ref{sect:implementation} and~\ref{sect:evaluation} describe the tool support we implemented for our approach, and the case studies we carried out to evaluate \acronym, respectively. Finally, we discuss related work in Section~\ref{sect:related}, and we conclude with a brief summary and we suggest directions for future work in Section~\ref{sect:conclusion}.

\section{Preliminaries \label{sect:preliminaries}}

\subsection{Parametric Markov chains}

\begin{definition2}
A discrete-time Markov chain is a tuple 
\begin{equation}
\label{eq:mc}
  M=(S,s_0,\mathbf{P},L),
\end{equation}
where: $S$ is a finite set of states; $s_0\!\in\! S$ is the initial state; $\mathbf{P}:S\!\times\! S\rightarrow [0,1]$ is a transition probability matrix such that, for any states $s,s'\!\in\! S$, $\mathbf{P}(s,s')$ is the probability of transition from $s$ to $s'$ and, for any $s\!\in\! S$, $\sum_{s'\in S} \mathbf{P}(s,s')\!=\!1$; and $L:S\rightarrow 2^\mathit{AP}$ is a function that maps every state $s\!\in\! S$ to those elements of an atomic proposition set $\mathit{AP}$ that hold in state $s$. 
\end{definition2}

\revised{To extend the range of properties that can be verified using discrete-time Markov chains, their states and/or transitions are often annotated with non-negative quantities termed \emph{rewards}.}

\begin{definition2}
A \emph{reward structure} over a \revised{discrete-time} Markov chain $M$ is a pair of functions $(\rho, \iota)$ that map the states and state transitions of $M$ to non-negative quantities called \emph{rewards}: $\rho:S\rightarrow [0,\infty)$ and $\iota:S\times S\rightarrow [0,\infty)$.
\end{definition2}

For Markov chains used in software performance and dependability engineering, the states may  correspond to different SUV configurations, to different operations being executed, to different outcomes of these executions, etc. In these models, the rewards may specify expected execution times, resource use and other quantitative characteristics  of the operations carried out by the SUV. Finally, the continuous variables used to define the unknown transition probabilities and rewards represent parameters of the SUV components.

\begin{definition2} 
A \emph{parametric \revised{(discrete-time)} Markov chain} is a discrete-time Markov chain \revised{comprising one or several unknown} state transition probabilities and/or rewards \revised{that are} specified as rational functions (i.e., as fractions whose numerators and denominators are polynomial functions, e.g., $1-p$ or $(1-p_1)/p_2$~\cite{maths-encyclopedia}) over a set of continuous variables~\cite{Daws:2004:SPM:2102873.2102899}.
\end{definition2}

\revised{The continuous variables used to specify the transition probabilities and/or rewards of parametric Markov chains correspond to parameters of the SUV.}

\subsection{Probabilistic computation tree logic}

To verify the nonfunctional requirements of a system modelled by a parametric Markov chain, these requirements are expressed in rewards-extended \cite{Andova2004} probabilistic computation tree logic (PCTL) \cite{Ciesinski2004,Hansson1994} with the syntax:
\begin{equation}
\label{eq:pctl}
\begin{array}{l}
\Phi::=  true \,\vert\, a \,\vert\, \Phi \wedge \Phi \,\vert\, \neg \Phi \,\vert\,  \mathcal{P}_{\bowtie p} [\Psi]\\
\Psi::= \mathrm{X}\; \Phi \;\vert\; \Phi\; \mathrm{U}^{\leq k}\, \Phi \;\vert\; \Phi\; \mathrm{U}\; \Phi\\
\Theta::=  \mathcal{R}_{\bowtie r} [\mathrm{I}^{=k}] \,\vert\,
           \mathcal{R}_{\bowtie r} [\mathrm{C}^{\leq k}] \,\vert\,
           \mathcal{R}_{\bowtie r} [\mathrm{F}\, \Phi] \,\vert\,
           \mathcal{R}_{\bowtie r} [\mathrm{S}]
\end{array}
\end{equation}
where  $\Phi$ is a \emph{state formula}, $\Psi$ a \emph{path formula}, $\Theta$ a \emph{reward state formula}, $k\!\in\! \mathbb{N}$ a timestep bound, $\bowtie\:\in\!\{ \mathord{\leq},<, \mathord{\geq},>\}$ a relational operator, $p\in[0,1]$ a probability bound, $r\geq 0$ a reward bound, and $a\!\in\! AP$ an atomic proposition. 

The PCTL semantics is defined using a satisfaction relation $\models$. Given a state $s$ of a Markov chain $M$, $s\models \Phi$ means ``$\Phi$ holds in state $s$'', and we have: always $s\models true$; $s \models a$ iff $a\in L(s)$; $s \models \neg \Phi$ iff $\neg (s\models \Phi)$; $s\models \Phi_1 \wedge \Phi_2$ iff $s\models \Phi_1$ and $s\models \Phi_2$; and $s\models \mathcal{P}_{\bowtie p} [\Psi]$ iff the probability $x$ that paths starting at state $s$ (i.e., sequence of states $s_1s_2s_3\ldots$ such that $s_1=s$ and $\forall i>0:\mathbf{P}(s_{i},s_{i+1})>0$) satisfy the path property $\Psi$ satisfies $x \bowtie p$. The \emph{next formula} $\mathrm{X}\;\Phi$ holds for a path if $\Phi$ is satisfied in the next state on the path; the \emph{bounded until formula} $\Phi_1\, \mathrm{U}^{\leq k}\, \Phi_2$ holds for a path iff $\Phi_1$ holds in the first $i<k$ path states and $\Phi_2$ holds in the $(i+1)$-th path state; and the \emph{unbounded until formula} $\Phi_1\,\mathrm{U}\, \Phi_2$ removes the bound $k$ from the  time-bounded until formula. 
Finally, the four reward state formulae $\Theta$ from~\eqref{eq:pctl} use the reward operator $\mathcal{R}$ to verify whether the expected reward $x$: at timestep $k$; accumulated up to timestep $k$; accumulated to reach a state that satisfies $\Phi$; and at steady state, respectively, satisfies $x\bowtie r$.
For a detailed description of the PCTL semantics, see \cite{Andova2004,Ciesinski2004,Hansson1994}. 

Additionally, the notation $\mathrm{F}^{\leq k}\:\Phi \equiv \mathsf{true}\:\mathrm{U}^{\leq k}\:\Phi$ and $\mathrm{F}\:\Phi \equiv \mathsf{true}\:\mathrm{U}\:\Phi$ is used when the left-hand side of a bounded until and until formula, respectively, is $\mathit{true}$; and $\mathcal{P}_{\!=?} [\:\cdot\:]$ and $\mathcal{R}_{\!=?} [\:\cdot\:]$ are used to denote the value of the probability and expected reward from a PCTL state and reward state formula, respectively.

\subsection{Formal verification with confidence intervals \label{sect:fact}}

\emph{Formal verification with confidence intervals}~\cite{calinescu2015formal} is a mathematically based technique for the computation of confidence intervals for nonfunctional properties of systems with stochastic behaviour. Given a parametric Markov chain $M=(S,s_0,\mathbf{P},L)$ that models the behaviour of a SUV, a PCTL formula $\mathcal{P}_{\!=?} [\:\cdot\:]$ or $\mathcal{R}_{\!=?} [\:\cdot\:]$ corresponding to a nonfunctional property of the SUV, and a confidence level $\alpha\in(0,1)$, the technique computes an $\alpha$ confidence interval for the property. 

To perform this computation, the technique \rerevised{starts by calculating confidence intervals for the parameters of the Markov chain by using} observations of the outgoing transitions from all states with unknown outgoing transition probabilities. 
Assuming that $Z\subseteq S$ is the subset of these states, the required observations are provided by a function 
\begin{equation}
    \label{eq:observations}
    O:Z\times S\rightarrow \mathbb{N} 
\end{equation}
that maps each pair of states $(z,s)\in Z\times S$ to the number $O(z,s)$ of 
\rerevised{times the transition} from $z$ to $s$ \rerevised{has been observed, within the given observation time}. 
\revised{We note that $O(z,s)=0$ for the states $s\in S$ for which no transition from $z$ to $s$ was observed. This may be the case even when such transitions are possible (e.g., because they correspond to rare SUV events), and the potential lack of these observations is the very reason for using confidence intervals instead of point estimates in the formal verification.}
Given such observations, the confidence interval computation is done in three stages. In the first stage, a confidence interval is calculated for each parameter of the Markov chain. In the next stage, parametric model checking is used to obtain a closed-form expression \revised{(i.e., a mathematical expression containing only the basic arithmetic operators $+$, $-$, $\times$ and $/$, and exponent, e.g., $\frac{p_1(1-p_2)^2}{1-p_3p_4}$)} for the nonfunctional property.\footnote{Parametric model checking is supported by a growing number of model checkers, including PARAM~\cite{Hahn2010}, PRISM~\cite{prism}, Storm~\cite{Dehnert2017} and ePMC/fPMC~\cite{calinescu2019efficient,FangCGA2021}.} This expression is \rerevised{a} rational function over the SUV parameters, and is a byproduct of the technique exploited by our \acronym\ approach. Finally, in the third stage, the parameter confidence intervals and the property expression are used to establish the confidence interval for the nonfunctional property of interest. For a detailed description of the technique and of a model checker that implements it see~\cite{calinescu2015formal} and~\cite{calinescu2016fact}, respectively.  \revised{This model checker supports non-nested PCTL properties $\mathcal{P}_{\!=?} [\:\cdot\:]$, and all types of PCTL reward properties.}

The number of available observations and the confidence level $\alpha$ used by the technique influence the width of the confidence interval. Thus, few observations and large $\alpha$ values yield wide confidence intervals that may contain the lower/upper bound that a  nonfunctional requirement specifies for the property. In this case, verifying whether the requirement is satisfied or not is impossible, and additional observations need to be collected to allow the computation of a narrower confidence interval that does not contain the bound.

\section{Motivating example \label{sect:example}}

To motivate our \acronym\ approach, we use a tele-assistance service-based system (TAS) introduced in  \cite{baresi2007validation}. TAS aims to support a patient suffering from a chronic condition in the comfort of their home by using: (i)~a set of vital-sign monitoring sensors mounted on a medical device worn by the patient; and (ii)~remote assistance services offered by emergency, medical and pharmacy service providers. 
 
\begin{figure}
    \centering 
    \includegraphics[width=\hsize]{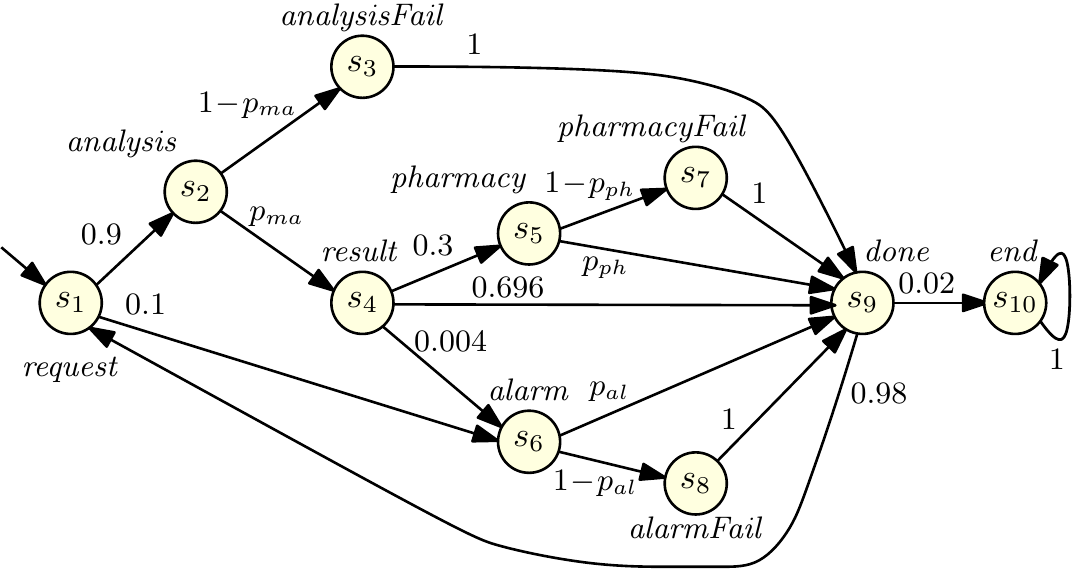}
    
    \vspace*{-1mm}
    \caption{Parametric Markov chain modelling the TAS workflow (adapted from~\cite{6693145})} \label{fig:TAS-model}
\end{figure}

\revised{The workflow implemented by the TAS system is modelled by the parametric Markov model from Figure~}\ref{fig:TAS-model}.\footnote{As described in~\cite{calinescu2013using,ghezzi2013managing} \rerevised{and summarised in Appendix~A}, discrete-time Markov models for software systems can be derived from established software engineering models such as UML activity diagrams.} Periodically, the patient's vital signs are measured by the wearable device \revised{(a workflow step whose completion is modelled by the Markov model transition from the initial state $s_1$ to state $s_2$)}, and a third-party medical analysis service is invoked to analyse them in conjunction with the patient's medical record \revised{(state $s_2$). This invocation may succeed (transition $s_2\rightarrow s_4$) or fail (transition $s_2\rightarrow s_3$)}. Depending on the results of this analysis \revised{(state $s_4$)}, TAS may confirm that the patient is fine \revised{(transition $s_4\rightarrow s_9$)}, may invoke a pharmacy service to request the delivery of different medication to the patient's home \revised{(state $s_5$, followed by a transition to $s_9$ if the invocation succeeds or to $s_7$ otherwise)}, or may invoke an alarm service \revised{(state $s_6$, followed by a transition to $s_9$ if the invocation succeeds or to $s_8$ otherwise)}. The invocation of the alarm service is also triggered when the patient presses a panic button on the wearable device \revised{(modelled by the transition $s_1\rightarrow s_6$)}, and results in a medical team being dispatched to provide emergency assistance to the patient. \revised{Once the TAS system is done with executing this workflow (state $s_9$), it may return to the state in which it awaits further requests (state $s_1$) or it may reach the end of its deployment (state $s_{10}$, whose self-loop of probability $1$ shows that no further transition to other Markov model states is possible).} We assume that the operational profile of the  system is known (e.g., from previous deployments) and is given by the probabilities annotating the \revised{outgoing transitions from states $s_1$, $s_4$ and $s_9$ from Figure~}\ref{fig:TAS-model}\revised{.}

\begin{table*}
	\renewcommand{\arraystretch}{1.2}
	\centering
	\caption{Nonfunctional requirements for the TAS system}
	
    \vspace*{1.5mm}
    \begin{sffamily}
    \def\tabcolsep{3.75pt}
	\begin{footnotesize}
		\begin{tabular}{p{0.4cm} p{8.8cm} p{5.3cm}} %
			\toprule
			\textbf{ID} 
			&\textbf{Requirement} & \textbf{PCTL formula}
			\\
			\midrule
			R1
			& The probability that an alarm failure ever occurs during the lifetime of the TAS system shall be below $0.26$. & $\mathcal{P}_{<0.26}[\mathrm{F}\;\mathit{alarmFail}]$\\
			R2
			& The probability that the handling of a request by the TAS workflow ends with a service failure shall be below $0.04$. & $\mathcal{P}_{<0.04}[\neg \mathit{done}\;\mathrm{U}\; \mathit{serviceFail}]$\\
			R3
			& The probability that an invocation of the medical analysis service is followed by an alarm failure shall be below $0.0003$.
			& $\mathcal{P}_{<0.0003}[\neg \mathit{done}\;\mathrm{U}\;\mathit{alarmFail}\{\mathit{analysis}\}]$\revised{$^\dagger$}\\
			\bottomrule
			\multicolumn{3}{l}{\revised{$^\dagger$We adopt the extended PCTL syntax of the PRISM model checker~}\cite{prism}\revised{ to specify that this PCTL formula should be}}\\
			\multicolumn{3}{l}{\revised{verified for the Markov chain state for which the atomic proposition $\mathit{analysis}$ holds (i.e., state $s_2$) instead of the initial}}\\
			\multicolumn{3}{l}{\revised{ state $s_0$, which is the default.}}
		\end{tabular}
	\end{footnotesize}
    \end{sffamily}
	\label{table:TAS-requirements}	
	
	\vspace*{-1.5mm}
\end{table*}

We suppose that a team of software engineers wants to verify whether the third-party services they consider for the implementation of the TAS system satisfy---at 95\% confidence level---the nonfunctional requirements from 
Table~\ref{table:TAS-requirements}. We assume that the three services are yet to be tested, and therefore the success probabilities $p_\mathit{ma}$, $p_\mathit{ph}$ and $p_\mathit{al}$ for the medical analysis service, pharmacy service and alarm service, respectively, are unknown \revised{continuous variables, as specified in Definition~3}. As such, the Markov chain that the engineers can use to verify the TAS requirements is parametric (Figure~\ref{fig:TAS-model}), and the three services must be tested to observe how many of their executions succeed and how many fail (e.g., by not finishing in a timely manner). With these observations, the engineers can use formal verification with confidence intervals (cf.~Section~\ref{sect:fact}) to compute 95\% confidence intervals for the probabilities from the three TAS requirements, which are formally expressed in PCTL in the last column from Table~\ref{table:TAS-requirements}. Furthermore, once enough observations are available, these confidence intervals will be sufficiently narrow to ensure that the bounds from the requirements in Table~\ref{table:TAS-requirements} (i.e., $0.26$ for requirement \textsf{R1}, $0.04$ for \textsf{R2}, and $0.0003$ for  \textsf{R3}) fall outside the intervals, allowing the engineers to verify whether the requirements are satisfied or not.

However, under the realistic assumption that service invocations take non-negligible time, the engineers will want to complete this verification with as few invocations (i.e., observations) of each service as possible. Minimising this testing effort is particularly important when the verification needs to be performed at runtime, e.g., to find a suitable replacement for a failed component of a system, or when testing a system component has some other cost associated with it (e.g., an invocation charge paid to the provider of a service, or using battery energy on an embedded system). Deciding how many observations to obtain for each service in order to complete the verification of the requirements with minimal testing effort is very challenging. Our \acronym\ verification approach addresses this challenge as described in the next section.

\section{The \acronym\ verification approach \label{sect:approach}}

\subsection{Problem definition \label{subsect:problem}}

The \acronym\ verification approach is applicable to systems comprising $m>1$ components that can be tested independently. We consider a component-based system whose $n\geq 1$ nonfunctional requirements are of the form 
\begin{equation}
  \label{eq:reqs}
  \mathit{prop}_i\bowtie_i \mathit{bound}_i,
\end{equation}
where, for all $i\in\{1,2,\ldots,n\}$, $\mathit{prop}_i$ is a nonfunctional system property (e.g., reliability or\rerevised{, through the use of properties that compute the expected reward accumulated to reach states that model the completion of operations,} response time), $\bowtie_i\:\in\!\{<, \mathord{\leq}, \mathord{\geq},>\}$, $\mathit{bound}_i\in\mathbb{R}$ places a constraint on the acceptable values of $\mathit{prop}_i$, and the $i$-th requirement can be expressed as a PCTL formula $\mathcal{P}_{\bowtie_i \mathit{bound}_i}[\:\cdot\:]$ or $\mathcal{R}_{\bowtie_i \mathit{bound}_i}[\:\cdot\:]$ over a parametric Markov chain $M=(S,s_0,\mathbf{P},L)$. Given such a system, the verification problem addressed by \acronym\ is to verify whether the $n$ nonfunctional requirements~\eqref{eq:reqs} are satisfied at confidence level $\alpha\in(0,1)$:
\begin{enumerate}
    \item with \rerevised{minimum} overall testing cost;
    \item by using a (possibly empty) initial set of observations given by an observation function $O_0:Z\times S\rightarrow \mathbb{N}$ with the semantics from \eqref{eq:observations};
    \item by obtaining additional observations through unit-testing the $m$ system components as required, where each unit test of the $j$-th component:
        (i)~generates one additional observation of an outgoing transition for every state in a non-empty set $Z_j\subset Z$, such that the state sets $Z_1$, $Z_2$, \ldots, $Z_m$ are disjoint and $\bigcup_{j=1}^m Z_j = Z$;
        and (ii)~has an associated cost $\mathit{cost}_j$, that may represent testing time, resources, price, or a combination thereof.
\end{enumerate}
Using the notation $[l_i,u_i]$ to denote the $\alpha$-confidence interval that can be computed for $\mathit{prop}_i$ from~\eqref{eq:reqs} after obtaining $n_1, n_2, \ldots, n_m$ additional observations for component $1, 2, \ldots, m$, respectively, the  problem addressed by \acronym\ is to find $n_1, n_2, \ldots, n_m\geq 0$ such that the overall testing cost
\begin{equation}
  \sum_{j=1}^m n_j\mathit{cost}_j \textrm{ is minimised}
\end{equation}
subject to
\begin{equation}
\label{eq:all-reqs-satisfied}
\begin{array}{l}
  \forall i=1..n:(\bowtie_i\in\{<, \mathord{\leq}\} \land u_i\bowtie_i\mathit{bound}_i)\\
  \qquad\qquad\qquad\qquad\lor\;(\bowtie_i\in\{>, \mathord{\geq}\}\land l_i\bowtie_i\mathit{bound}_i)
\end{array}
\end{equation}
or
\begin{equation}
\label{eq:exists-req-violated}
\!\!\begin{array}{l}
  \exists i=1..n:(\bowtie_i=< \land\: l_i\!\geq\!\mathit{bound}_i)\lor (\bowtie_i= \mathord{\leq} \land\: l_i\!>\!\mathit{bound}_i)\\
  \qquad\quad\:\lor\;(\bowtie_i=>\land\: u_i\!\leq\! \mathit{bound}_i)\lor(\bowtie_i=\geq\land\: u_i\!<\!\mathit{bound}_i).
\end{array}
\end{equation}
The constraints \eqref{eq:all-reqs-satisfied} and~\eqref{eq:exists-req-violated} correspond to the scenarios where all requirements~\eqref{eq:reqs} are satisfied and where at least one of requirements~\eqref{eq:reqs} is violated (meaning that the set of requirements as a whole is violated), respectively.

Due to the epistemic uncertainty associated with this verification problem and the stochastic nature of the com\-po\-nent-testing results, a \revised{strategy guaranteed to achieve a} minimum overall testing cost \revised{does not exist.} \rerevised{We illustrate this limitation with an example. Consider a system that uses two web services, A and B. This system implements a simple workflow: first, it invokes service A, which is available with probability $p_A$; next, it invokes service B, which is available with probability $p_B$; next, it stops. Suppose that we want to establish whether the probability of successful invocation of both services is at least $0.9$ (i.e., whether $p_Ap_B\geq 0.9$) at confidence level $\alpha=0.95$. If the two unknown probabilities are $p_A=0$ (i.e., service A is never available) and $p_B=0.95$, then allocating all the testing effort to unit-testing service A is the cheapest strategy for establishing that the requirement is violated, as this strategy will quickly show that $p_A$ cannot be large enough for the requirement to be satisfied. Conversely, if $p_A=0.95$ and $p_B=0$, unit-testing only service B is the cheapest strategy. However, with no prior knowledge about $p_A$ and $p_B$, it is impossible to always choose the best testing strategy.} Therefore, our objective is to achieve an overall testing cost that is, on average, significantly lower than the cost associated with uniformly testing all components. 
\revised{Furthermore, for practical reasons, we add the constraint that the overall cost does} not exceed a predefined testing budget $\mathit{budget}\in \mathbb{N}$.

\begin{example}
Consider the TAS system from our motivating example. Its $n=3$ nonfunctional requirements \textsf{R1--R3} from Table~\ref{table:TAS-requirements} are of the form in~\eqref{eq:reqs}, are expressed as PCTL formulae over the parametric Markov chain from Figure~\ref{fig:TAS-model}, and need to be verified at confidence level $\alpha=0.95$. The set of Markov chain states with unknown outgoing transition probabilities is $Z=\{s_2,s_5,s_6\}$, and the (empty) initial set of observations is defined by the function $O_0(z,s)=0$ for any $(z,s)\in Z\times S$. The system comprises $m=3$ components that can be tested independently: the medical analysis service (component~1), the pharmacy service (component~2) and the alarm service (component~3). Additionally, invoking one of these services once provides an additional observation of an outgoing transition for the state in one of the disjoint sets $Z_1=\{s_2\}$, $Z_2=\{s_5\}$ and $Z_3=\{s_6\}$, where $Z_1\cup Z_2\cup Z_3=Z$. Finally, to fully cast the task of verifying requirements \textsf{R1--R3} in the format from our problem definition, we assume that a testing budget $\mathit{budget}\!=\!150000$ is available to complete the verification, and that the per-invocation costs of testing the three services are $\mathit{cost}_1\!=\!1$, $\mathit{cost}_2\!=\!1$ and $\mathit{cost}_3\!=\!2$, e.g., based on the ratios between their mean execution times. We chose a very large budget to ensure that the verification completes without exhausting the budget; this may be unacceptable in practice, in which case the verification might terminate without a decisive result (e.g., if the actual values of some of the $n$ nonfunctional properties are extremely close to their associated bounds).
\end{example}

\subsection{\acronym\ verification process}

\begin{figure*}
    \centering 
    \includegraphics[width=0.7\hsize]{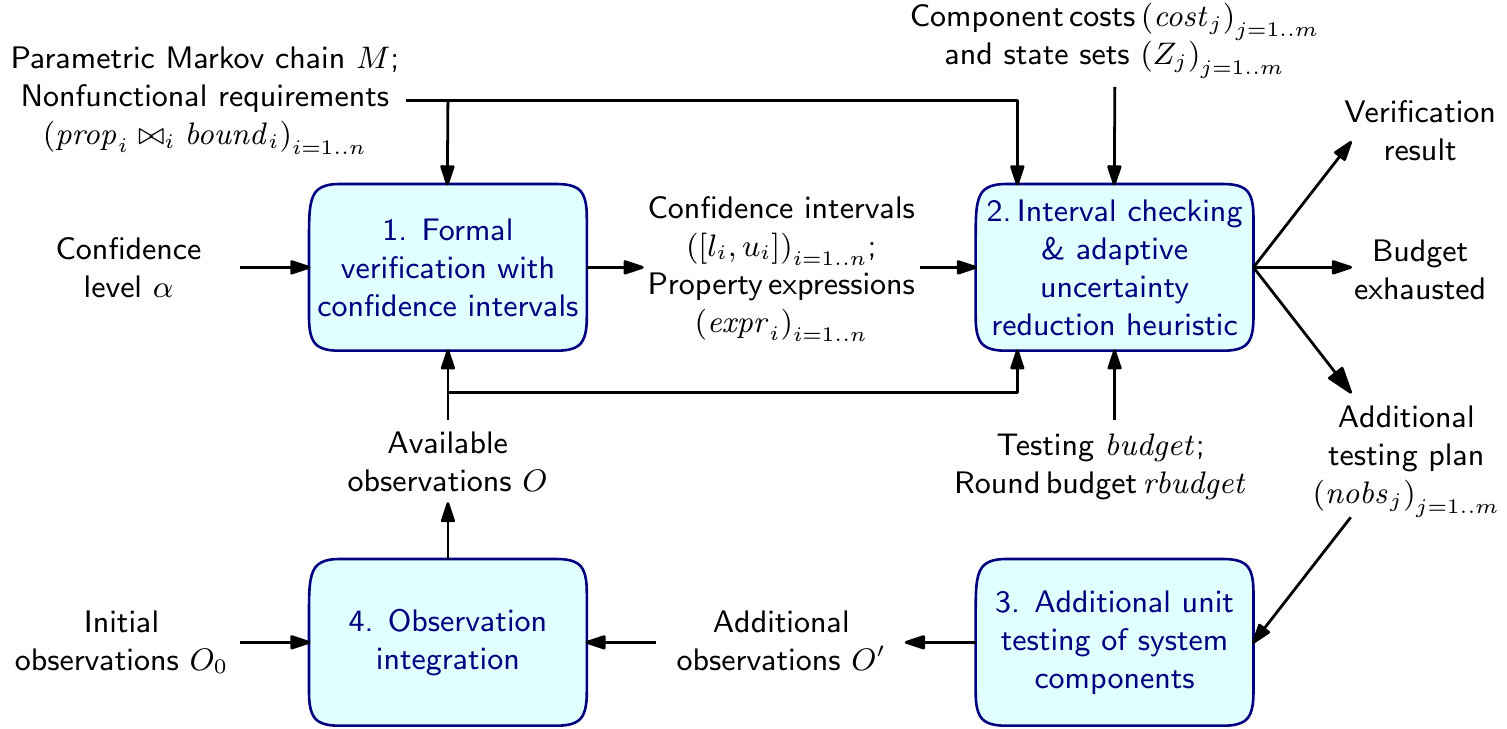}
    \caption{Iterative four-step process of the \acronym\ quantitative verification with adaptive uncertainty reduction} \label{fig:approach}
\end{figure*}

To solve the problem from Section~\ref{subsect:problem}, \acronym\ employs the iterative verification process depicted in Figure~\ref{fig:approach}. Each \emph{round} (i.e., iteration) of this process comprises the four steps described below. 

In the first step, formal verification with confidence intervals~\cite{calinescu2015formal} is used to compute confidence intervals $[l_1,u_1]$, $[l_2,u_2]$, \ldots, $[l_n,u_n]$ at confidence level $\alpha$ and (as a byproduct, as explained in Section~\ref{sect:fact}) closed-form expressions $\mathit{expr}_1$, $\mathit{expr}_2$, \ldots, $\mathit{expr}_n$ for the $n$ properties from the nonfunctional requirements~\eqref{eq:reqs}. The observations $O$ used to compute the $n$ confidence intervals include the initial observations $O_0$ and, starting with the second iteration, all the additional observations $O'$ obtained in the previous iterations of the process. If the observation set $O_0$ is empty, then the confidence interval $[l_i,u_i]$ computed for the $i$-th property in the first iteration is $[0,1]$ or $[0,\infty)$, depending on whether the $i$-th requirement is of the form $\mathcal{P}_{\bowtie_i \mathit{bound}_i}[\:\cdot\:]$ or $\mathcal{R}_{\bowtie_i \mathit{bound}_i}[\:\cdot\:]$.

\revised{In the second step, VERACITY checks whether the $n$ confidence intervals are sufficiently narrow for either constraint~}\eqref{eq:all-reqs-satisfied}\revised{ or constraint~}\eqref{eq:exists-req-violated}\revised{ to hold. If either of these conditions are met, the verification problem is solved, and a verification result is produced. Otherwise}, 
    additional observations are needed to complete the verification. Two cases are possible in this situation. First,  
    when the testing $\mathit{budget}$ was fully utilised in the previous \acronym\ rounds, the process terminates with an inconclusive `budget exhausted' result. Otherwise, testing budget is still available, and the \acronym\ adaptive uncertainty reduction heuristic detailed in Section~\ref{subsect:heuristic} is used to calculate the numbers of additional component observations $\mathit{nobs}_1$, $\mathit{nobs}_2$, \ldots, $\mathit{nobs}_m$ for the next round of the verification process, where 
    \begin{equation}
        \label{eq:round-budget-partition}
        \sum_{j=1}^m \mathit{nobs}_j\mathit{cost}_j\approx \mathit{rbudget}
    \end{equation} 
    and $\mathit{rbudget}\in[0,\mathit{budget}]$ is a parameter of the \acronym\ approach called the \emph{round budget}.\footnote{Equality cannot be always achieved in~\eqref{eq:round-budget-partition} because $\mathit{nobs}_1$, $\mathit{nobs}_2$, \ldots, $\mathit{nobs}_m$ must take non-negative integer values.} The heuristic is adaptive in the sense that these numbers of additional observations vary from round to round, as the heuristic takes into account the actual observations from all the previous rounds (and any initial observations that may be available). The maximum number of rounds for the verification process is $\lceil \mathit{budget}/\mathit{rbudget} \rceil$. Accordingly, larger round budgets yield fewer rounds, and therefore less opportunity for adaptation but lower overheads (due to the fewer rounds). In contrast, smaller round budgets lead to more rounds, which offer more opportunity for adaptation but  also come with higher overheads.

In the third step of the \acronym\ process, $\mathit{nobs}_j$ tests of component $j$ are carried out for $j=1..m$. As we will explain in Section~\ref{sect:implementation}, these tests can be fully automated, or can be performed by a software engineer when requested by the \acronym\ verification tool. The results of these tests are then encoded as an observation function $O'$ with the format from~\eqref{eq:observations}.

Finally, in the fourth and last step of \acronym, the new observations $O'$ are integrated with all the observations obtained in the previous rounds of the process and the initial observations $O_0$, and the combined set of all available observations $O$ is used in the next round of the verification process.

\begin{example}
For the three requirements for the TAS system from our motivating example, $\bowtie_1=\:\bowtie_2=\:\bowtie_3=<$. Accordingly, the requirements will be verified as satisfied at \revised{95}\% confidence level if the observations acquired over successive rounds of the \acronym\ verification process (and within the available $\mathit{budget}$) lead to the calculation of \revised{95}\% confidence intervals $\left([l_i,u_i]\right)_{i=1..3}$ that satisfy $u_1<\mathit{bound}_1$, $u_2<\mathit{bound}_2$ and $u_3<\mathit{bound}_3$. If, on the other hand, $l_i\geq\mathit{bound}_i$ for any $i\in\{1,2,3\}$ in one of the verification rounds, the verification process will decide that requirement~$i$ is violated at \revised{95}\% confidence level, and will terminate in that round. Finally, if the testing $\mathit{budget}$ is used up before sufficient observations of the medical analysis, pharmacy and alarm services are obtained to allow either of these decisions to be made, the verification process will terminate with a `budget exhausted' outcome.
\end{example}

\begin{algorithm*}[t]
\caption{Adaptive uncertainty reduction heuristic} \label{algorithm:heuristic}
	\begin{algorithmic}[1]
		\Function{\hspace*{-0.5mm}NewObs}{$\mathit{rbudget}$, 
		\hspace*{-0.5mm}$M$,  \hspace*{-0.5mm}$(\mathit{prop}_i\!\bowtie_i\!\mathit{bound}_i)_{i=1..n}$, \hspace*{-0.5mm}$([l_i,u_i])_{i=1..n}$, 
		\hspace*{-0.5mm}$(\!\mathit{expr}_i)_{i=1..n}$, 
		\hspace*{-0.5mm}$(\!\mathit{cost}_j)_{j=1..m}$, 
		\hspace*{-0.5mm}$(Z_j)_{j=1..m}$, 
		\hspace*{-0.5mm}$O$, 
		\hspace*{-0.5mm}\revised{$\epsilon_1$,} 
		\hspace*{-0.5mm}\revised{$\epsilon_2$}}
		\State $\mathit{U}=\{i\in 1..n \mid \mathit{bound}_i\in[l_i,u_i] \}$ \Comment{\textsf{desideratum D1}} \label{l:desideratum1}
		\If{$\exists i\in \mathit{U}: |\mathit{bound}_i\!-\!\textsc{Wrong}(\bowtie_i,l_i,u_i)| / |\mathit{bound}_i\!-\!\textsc{Right}(\bowtie_i,l_i,u_i)|<\epsilon_1$} \Comment{\textsf{desideratum D2}} \label{l:desideratum2}
		    \State $\mathit{R}\gets \left\{\mathsf{argmin}_{i\in \mathit{U}}|\mathit{bound}_i\!-\!\textsc{Wrong}(\bowtie_i,l_i,u_i)| / |\mathit{bound}_i\!-\!\textsc{Right}(\bowtie_i,l_i,u_i)|\right\}$ \label{l:R-violated}
		\Else
		    \State $\mathit{R}\gets\mathit{U}$ \label{l:R-not-violated}
		\EndIf \label{l:desideratum2-End}
		
		\State $\mathit{paramEstimate} \gets \textsc{EstimateParams}(M,O)$ \label{l:estimate}
		\State $(\mathit{relevance}_j \gets 0)_{j=1..m}$ \label{l:initialise}
		\For{$i\in \mathit{R}$} \label{l:outerloop}
		    \State $\mathit{weight} = (u_i-l_i)/\max\{|\mathit{bound}_i-(l_i+u_i)/2|, \epsilon_2\}$ \Comment{\textsf{desideratum D3}} \label{l:desideratum3}
		    \For{$j=1$ \textbf{to} $m$}  \label{l:innerloop}
		        \State $\mathit{sensitivity}\gets \sum_{p\in \textsc{Params}(M,\mathit{Z}_j)} \left|\frac{\partial \mathit{expr}_i(\mathit{paramEstimate})}{\partial p}\right|$ \Comment{\textsf{desideratum D4}} \label{l:desideratum4} 
		        \State $\mathit{relevance}_j \gets  \mathit{relevance}_j +\mathit{weight}\cdot\mathit{sensitivity}$ \label{l:contribution}
		    \EndFor  \label{l:innerloopEnd}
		\EndFor \label{l:outerloopEnd}
		\For{$j=1$ \textbf{to} $m$} \label{l:nobsLoop}
			\State \revised{$\mathit{nobs}_j \gets \left\lfloor \mathit{rbudget}\cdot  \frac{\mathit{relevance}_j}{\sum_{k=1}^m\mathit{relevance}_k} \cdot\frac{1}{\mathit{cost}_j}\right\rfloor$} \Comment{\textsf{desideratum D5}} \label{l:nObsCalculation}
		\EndFor \label{l:nobsLoopEnd}
		\State \Return {$(\mathit{nobs})_{j=1..m}$}
	    \EndFunction
	\end{algorithmic}
\end{algorithm*}

\subsection{Adaptive uncertainty reduction heuristic \label{subsect:heuristic} }

\subsubsection{Desiderata \label{subsect:desiderata}}

Before describing the \acronym\ heuristic for partitioning the round testing budget $\mathit{rbudget}$ among the $m$ system components, we present a set of desirable properties (i.e., \emph{desiderata}) that we propose for any such heuristic:
\begin{enumerate}
    \item[D1.] The requirements already verified as satisfied in previous rounds should not influence the partition of the round budget, as the uncertainty associated with these requirements has already been lowered enough to complete their verification.
    \item[D2.] Reaching a resolution on requirements \revised{whose verification requires additional data} (i.e., on requirements with $\mathit{bound}_i\in[l_i,u_i]$) that are \emph{likely} to be violated should be prioritised when the round budget is partitioned. This desideratum captures the fact that verifying a requirement as violated ends the verification process immediately. Such requirements can be identified by noting that their $\mathit{bound}_i$ is very close to the ``wrong'' end of the confidence interval $[l_i,u_i]$. For instance, if $\mathit{bound}_i$ were much closer to $l_i$ \revised{than} to $u_i$, narrowing down the confidence interval $[l_i,u_i]$ even slightly has a good chance (but is, of course, not certain) of showing that requirement $i$ is violated, and of terminating the verification process.
    \item[D3.] 
    If several requirements \revised{whose verification requires additional data} influence the partition of $\mathit{rbudget}$, 
    those requirements whose $\mathit{bound}_i$ value is closer to the middle of the confidence interval $[l_i,u_i]$ should have a bigger influence.  
    This desideratum reflects the fact that the verification of such requirements is particularly affected by epistemic uncertainty, as a significant narrowing of their confidence intervals is likely to be needed in order to decide whether they are satisfied or violated.
    \item[D4.] The $\mathit{rbudget}$ fraction allocated to each component should reflect the sensitivity of the  properties $\mathit{prop}_1$ to $\mathit{prop}_n$ from~\eqref{eq:reqs} to the parameters of that component. For instance, if the closed-form expression for the $i$-th requirement is $\mathit{expr}_i=p_1+0.01p_2$, where $p_1$ and $p_2$ are probabilities associated with components $C_1$ and $C_2$, respectively, component $C_1$ should be allocated a larger $\mathit{rbudget}$ fraction than $C_2$ (all other factors being equal).
    \item[D5.] When the testing of different components is expected to yield similar uncertainty reductions, the round budget partition should prioritise the components with lower testing costs: 
    the value of a new observation of a component is given by the ratio between the (expected) reduction in uncertainty brought by the observation and the cost of that observation.
\end{enumerate}

\subsubsection{Algorithm}

The numbers of new component observations 
$\mathit{nobs}_1$, $\mathit{nobs}_2$, \ldots, $\mathit{nobs}_m$   
for each \acronym\ verification round are computed by function \textsc{NewObs} from Algorithm~\ref{algorithm:heuristic}. This function takes the following arguments (cf.~Figure~\ref{fig:approach}):
\begin{itemize}
    \item the round testing budget $\mathit{rbudget}$;
    \item the Markov chain $M$ and the requirements $(\mathit{prop}_i\bowtie_i\mathit{bound}_i)_{i=1..n}$;
    \item the property confidence intervals $([l_i,\!u_i])_{i=1..n}$ and expressions $(\!\mathit{expr}_i\!)_{i=1..n}$ obtained in the previous step of the round; 
    \item the component testing costs $(\mathit{cost}_j)_{j=1..m}$ and associated state sets with unknown transition probabilities $(Z_j)_{j=1..m}$;
    \item the observations $O$ available at the start of the round;
    \item \revised{the configuration parameters $\epsilon_1,\epsilon_2\in(0,1)$, whose role is described later in this section}.
\end{itemize}
The algorithm has three parts. In the first part (lines~\ref{l:desideratum1}--\ref{l:desideratum2-End}), it identifies the set of \emph{relevant requirements} $R$ that will influence the partitioning of the round budget. According to desideratum~D1, the set $U$ of requirements \revised{whose verification requires additional data} is computed in line~\ref{l:desideratum1}. Next, the if statement in lines~\ref{l:desideratum2}--\ref{l:desideratum2-End} checks if $\mathit{bound}_i$ of any requirement \revised{from $U$} is much closer (i.e., $1/\epsilon_1$ times closer) to the ``wrong'' end of the confidence interval $[l_i,u_i]$ than to the ``right'' end, where:\\[-5mm]
\begin{itemize}
    \item the ``wrong'' end of $[l_i,u_i]$ is the end beyond which requirement $i$ is violated, i.e., $l_i$ if $\bowtie_i\:\in\!\{<, \mathord{\leq}\}$, and $u_i$ otherwise;
    \item the helper functions \textsc{Wrong}, \textsc{Right} return the respective ends of $[l_i,u_i]$;
    \item $\epsilon_1\!\in\!(0,1)$ is a \acronym\ configuration parameter.
\end{itemize}
As explained in desideratum~D2, requirements with this property are likely to be violated. Therefore, if any such requirements exist, only the requirement most likely to be violated is retained in the relevant requirement set $R$ (line~\ref{l:R-violated}). Otherwise, $R$ is initialised to include all the requirements \revised{whose verification requires additional data}  (line~\ref{l:R-not-violated}).

The second part of the algorithm (lines~\ref{l:estimate}--\ref{l:outerloopEnd}) starts by using the observations $O$ to calculate estimates for each unknown transition probability (i.e., parameter) associated with a state from $Z=\bigcup_{j=1}^m Z_j$ (line~\ref{l:estimate}). This calculation is performed by the auxiliary function \textsc{EstimateParams}, which estimates the unknown transition probabilities between each state $z\in Z$ and each state $s\in S$ using the observed transition frequency $O(z,s)/\sum_{s'\in S} O(z,s')$. The special case when $\sum_{s'\in S} O(z,s')=0$ for one or more states $z\in Z$ may be encountered in the first round, as we allow an empty initial set of observations $O_0$ (cf.~Section~\ref{subsect:problem}). In this case, which we do not show in Algorithm~\ref{algorithm:heuristic} in order to keep the pseudocode simple, \textsc{EstimateParams} raises an exception and the round budget is split uniformly between the components whose state sets $Z_j$ include states with zero observations.

Next in this part of the algorithm, a component relevance measure $\mathit{relevance}_j$ is first initialised in line~\ref{l:initialise}, and then updated with contributions corresponding to the relevant requirements $R$ by the for loop from lines~\ref{l:outerloop}--\ref{l:outerloopEnd}. Each such contribution is the product of two factors (line~\ref{l:contribution}) that correspond to desiderata~D3 and~D4, respectively: 
\begin{itemize}
    \item $\mathit{weight}$, a factor calculated as the ratio between the width of the confidence interval $[l_i,u_i]$ and the distance between $\mathit{bound}_i$ and the middle of the interval $[l_i,u_i]$ (line~\ref{l:desideratum3}, where a small \acronym\ configuration parameter $0<\epsilon_2\ll 1$ is used to prevent a division by zero);
    \item $\mathit{sensitivity}$, a measure of the sensitivity of expression $\mathit{expr}_i$ to the epistemic uncertainty affecting the parameters of component $j$, calculated by summing the absolute value of the partial derivatives of $\mathit{expr}_i$ with respect to every parameter of component $j$ (line~\ref{l:desideratum4}),\revised{\footnote{\revised{Our $\mathit{sensitivity}$ measure resembles the Birnbaum's measure of component importance for multicomponent systems~}\cite{Birnbaum69}\revised{, which is often used in fault tree analysis~}\cite{barlow1975importance}\revised{.}}} where the set of all such parameters is provided by the auxiliary function \textsc{Params}, and the partial derivatives are evaluated for the parameter values estimated in line~\ref{l:estimate}.
\end{itemize}

The third and final part of the algorithm (lines~\ref{l:nobsLoop}--\ref{l:nobsLoopEnd}) decides the number of new observations for each component based on the relevance and testing cost of that component. \revised{In accordance with} desideratum D5, the number of new observations for component $j$ is calculated (in line~\ref{l:nObsCalculation}) by \revised{allocating to the component a fraction of $\frac{\mathit{relevance}_j}{\sum_{k=1}^m\mathit{relevance}_k}$ of $\mathit{rbudget}$, and dividing this ``component budget'' by $\mathit{cost}_j$.} 

To avoid ending with $\mathit{nobs}_1=\mathit{nobs}_2=$ $\ldots=\mathit{nobs}_m=0$, it is sufficient to use a round budget that satisfies the constraint $\mathit{rbudget}\geq \sum_{j=1}^m \mathit{cost}_j$ because:
\[
\begin{array}{l}
  \forall j=1..m:\mathit{nobs}_j=0\\ 
  \implies \forall j=1..m:\mathit{rbudget}\cdot\frac{\mathit{relevance}_j}{\sum_{k=1}^m \mathit{relevance}_k} < \mathit{cost}_j\\ \implies \sum_{j=1}^m \left( \mathit{rbudget}\cdot\frac{\mathit{relevance}_j}{\sum_{k=1}^m \mathit{relevance}_k}\right) < \sum_{j=1}^m \mathit{cost}_j\\
  \implies \mathit{rbudget} < \sum_{j=1}^m \mathit{cost}_j.
\end{array}
\]
To achieve good progress with the verification process, $\mathit{rbudget}$ should in fact be much larger (e.g., at least one order the magnitude larger) than $\sum_{j=1}^m \mathit{cost}_j$ in practice.

For improved readability, a couple of efficiency improvements are not included in Algorithm~\ref{algorithm:heuristic}. \revised{In particular}, the function \textsc{Params} and the partial derivatives required for factor $\mathit{sensitivity}$ (line~\ref{l:desideratum4}) can be precomputed once (in the first round of the \acronym\ verification process), as the SUV parameters associated with a component do not change; only the evaluations of the precomputed partial derivatives need to be done in each round, for the new $\mathit{paramEstimate}$ from that round. 

With these improvements in place, the complexity of algorithm is $\mathsf{O}(mn)$, because of the two nested for loops from lines~\ref{l:outerloop}--\ref{l:outerloopEnd} and  \ref{l:innerloop}--\ref{l:innerloopEnd}, respectively. This is typically negligible compared to the complexity of the formal verification with confidence intervals and the additional unit testing from steps one and three of the \acronym\ verification process, respectively.

\begin{example}
Figure~\ref{fig:example} shows the difference between the verification of the TAS nonfunctional requirements from Table~\ref{table:TAS-requirements} using the \acronym\ verification process from Figure~\ref{fig:approach} with: (a)~our adaptive uncertainty reduction heuristic from Algorithm~\ref{algorithm:heuristic}; and (b)~our heuristic replaced with a uniform splitting of the round testing budget among the three system components. These results were obtained assuming that the unknown probabilities from the Markov chain in Figure~\ref{fig:TAS-model} were $p_\mathit{al}=0.94$, $p_\mathit{ma}=0.99$ and $p_\mathit{ph}=0.95$, and using random number generators to synthesise additional observations $O'$ based on these probabilities in the additional unit testing step of the \acronym\ verification process from Figure~\ref{fig:approach}. The verification was performed with a round budget $\mathit{rbudget}=5000$, unlimited overall testing $\mathit{budget}$, and the default values $\epsilon_1=0.15$ and $\epsilon_2=10^{-6}$ for the two parameters of the \acronym\ heuristic from Algorithm~\ref{algorithm:heuristic}. 

    \begin{figure*}
        \centering 
        \includegraphics[trim=8mm 72mm 32mm 74mm, clip, width=0.7\hsize,]{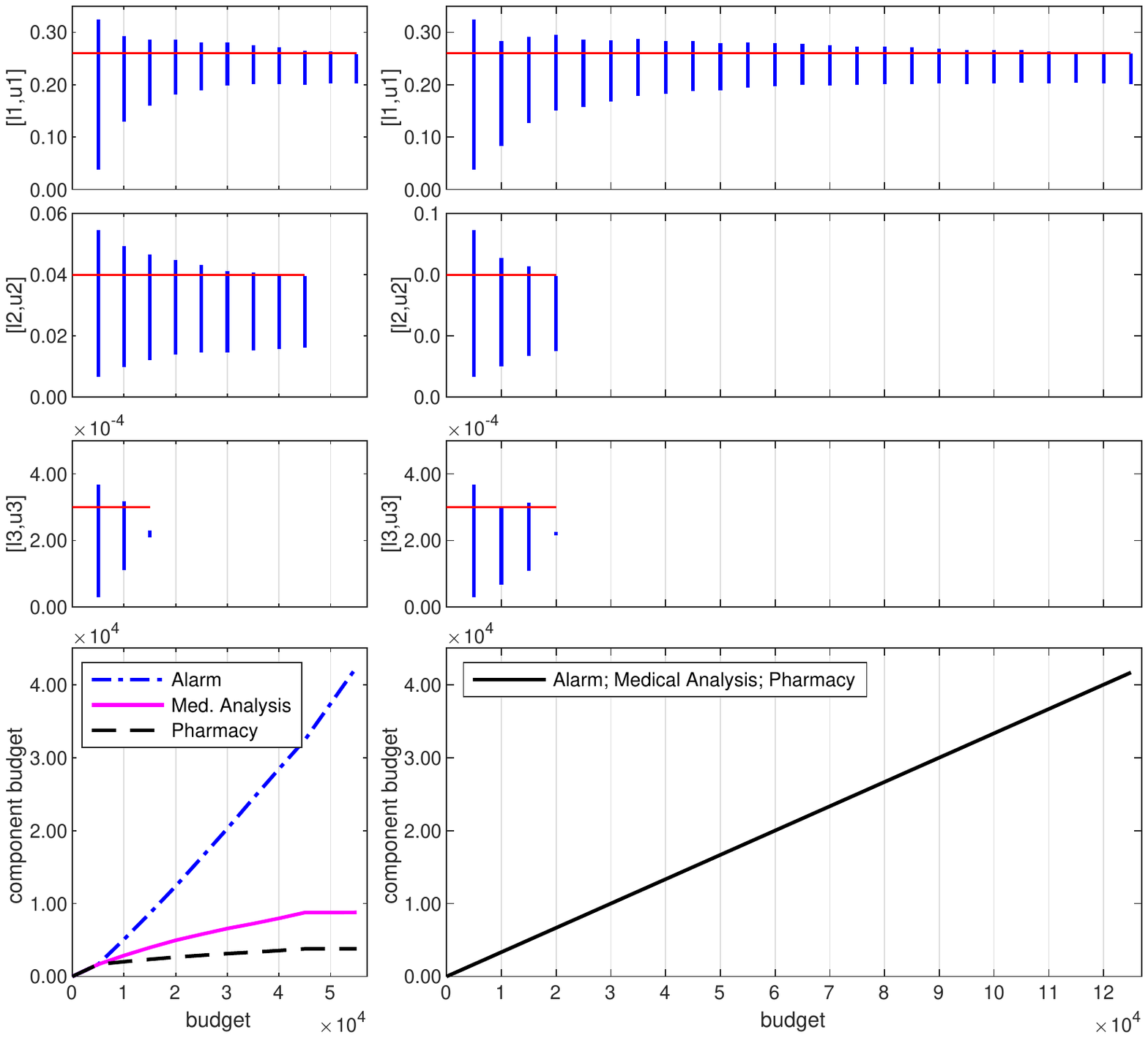}
        
        \begin{footnotesize}
        \hspace*{-5mm}(a) \acronym\ \hspace*{2.5cm} (b) uniform uncertainty reduction
        \end{footnotesize}
        
        \vspace*{-1mm}
        \caption{Verification of the nonfunctional requirements for the TAS system from the motivating example using (a)~\acronym\ adaptive vs.\ (b)~uniform uncertainty reduction} \label{fig:example}
        
        \vspace*{-2.5mm}
\end{figure*}

The top three pairs of graphs from Figure~\ref{fig:example} show the 95\% confidence intervals $([l_i,u_i])_{i=1..3}$ (depicted as vertical lines) for the nonfunctional properties from the three TAS requirements from Table~\ref{table:TAS-requirements}. These confidence intervals become narrower as additional observations are obtained in each round of the verification process, until they are narrow enough to fit completely under the $\mathit{bound}_i$ threshold (drawn as a horizontal line) from their associated requirement, i.e., until $u_i<\mathit{bound}_i$. As soon as this condition is met for a confidence interval $[l_i,u_i]$, that interval is no longer calculated in subsequent verification rounds. When the condition is met for all three confidence intervals, the epistemic uncertainty was reduced sufficiently to conclude that all requirements are satisfied, and the verification process terminates successfully. As shown by these graphs, \acronym\ and the uniform uncertainty reduction method finish the verification of each requirement after a different number of verification rounds, and \acronym\ completes the verification of the entire set of requirements with an overall testing cost of $55,000$ compared to a 127\% higher overall testing cost of $125,000$ for the approach based on uniform uncertainty reduction. We also note in these graphs that the lower bounds of confidence intervals are not always increasing \rerevised{and the upper bounds of confidence intervals are not always decreasing from one VERACITY iteration to the next}. As explained in~\cite{calinescu2015formal}\revised{, this is because the calculation of confidence intervals for the multinomially distributed transition probabilities (and therefore also for the properties) of Markov chains is conservative, so increasing the number of observations may occasionally widen the confidence interval slightly. This does not affect the validity of formal verification with confidence intervals.}

The bottom pair of graphs from Figure~\ref{fig:example} shows the cumulative testing cost per system component. When uniform uncertainty reduction is used, this cost is identical for all components, and is growing linearly with the number of verification rounds. In contrast, for the \acronym\ approach, the cumulative cost grows at different rates for different components. Furthermore, the rate of growth for any single component varies across verification rounds because \acronym\ continually \emph{adapts} its partitioning of the round budget to the observations acquired in previous rounds, and to the effect that these observations have on confidence intervals $([l_i,u_i])_{i=1..3}$.
\end{example}

\section{Implementation \label{sect:implementation}}

We implemented the \acronym\ verification process as a Java tool built on top of our FACT model checker~\cite{calinescu2016fact}. The \acronym\ tool takes as input: a parametric Markov chain $M$ expressed in the PRISM modelling language~\cite{prism} and annotated with the component costs $\left(\mathit{cost}_j\right)_{j=1..m}$ and state sets $\left(\mathit{Z}_j\right)_{j=1..m}$, and with the initial observations $O_0$; a set of PCTL-encoded nonfunctional requirements; and a confidence level $\alpha$. 

The overall testing $\mathit{budget}$ and round testing budget $\mathit{rbudget}$ are specified via a configuration file. In addition, this configuration file allows the user to optionally specify a component testing script that the tool can execute with the command\\[-2mm]
\[
    \texttt{testing-script}\;\: j\;\: nobs_j
\]
in order to obtain $\mathit{nobs}_j$ additional observations for component $j$ automatically in the third step of the \acronym\ verification process (cf.~Figure~\ref{fig:approach}). If provided, this script needs to run $\mathit{nobs}_j$ unit-test against component $j$ (e.g., by invoking the appropriate third-party service for the TAS system from our motivating example), and to return the $\mathit{nobs}_j$ observations from these tests as a list of numbers of transitions from the states in $Z_j$ to other states of the Markov chain $M$. Alternatively (i.e., if the testing script is not provided), the tool asks the user to supply the required $\mathit{nobs}_j$ observations interactively at each round of the verification process.

Our \acronym\ tool uses the model checkers FACT~\cite{calinescu2016fact} and PRISM~\cite{prism}, as well as  MATLAB\footnote{\url{http://www.mathworks.co.uk/products/matlab}} to compute the confidence intervals and property expressions in the first step of its verification process. As such, the tool supports the same fragment of PCTL as FACT, i.e., non-nested PCTL properties $\mathcal{P}_{\!=?} [\:\cdot\:]$ and all types of PCTL reward properties (cf.~Section~\ref{sect:fact}). The tool is freely available from our project website \url{https://www.cs.york.ac.uk/tasp/VERACITY}, together with detailed instructions and all the models, requirement sets, and results from this paper.

\section{Evaluation \label{sect:evaluation}}

We evaluated \acronym\ by performing an extensive set of experiments aimed at answering the following research questions (RQs).

\paragraph{RQ1} Does \acronym\ reduce the testing budget needed to verify a set of nonfunctional requirements compared to the baseline approach that partitions the testing budget of each verification round equally among SUV components?

\paragraph{RQ2} How effective is \acronym\ at reducing the testing budget in scenarios where the SUV components have different testing costs?

\paragraph{RQ3} What effect does adjusting the round budget have on the \acronym\ testing budget and verification time?

\vspace*{2.5mm}
\revised{Answering research questions RQ1 and RQ2 required the comparison of experimental data from the use of VERACITY and of the baseline approach. We carried out this comparison using established empirical methods from software engineering~}\cite{kampenes2007systematic,kitchenham2017robust,madeyski2018effect,madeyski2009test}\revised{ as follows. We started by applying the Shapiro-Wilk normality test, and established (as detailed in Sections~}\ref{subsect:RQ1}\revised{ and~}\ref{subsect:RQ2}\revised{) that our experimental data were not normally distributed. As such, we used three non-parametric methods to compare the two approaches:}
\begin{enumerate}
    \item \revised{We used the non-parametric Wilcoxon signed-rank test (as recommended, for instance, in~}\cite{kampenes2007systematic,madeyski2009test}\revised{) to assess whether VERACITY completes the verification with a smaller testing budget than the baseline approach.}
    \item \revised{We computed the \emph{probability of superiority} for depen\-dent-groups~}\cite{grissom2005effect}\revised{ to evaluate the effect size associated with the use of VERACITY---this is a robust non-parametric measure of effect size recommended for software engineering experiments~}\cite{kitchenham2017robust}\revised{; and we generated scatter plots for the visual inspection of the experimental data.}
    \item \revised{We calculated the difference in overall testing cost between VERACITY and the baseline approach, and we report the median value of these differences (where a negative value indicates a reduction in cost). Additionally, we plotted box plots (enabling the easy visual examination) of the percentage difference between the testing budgets required by the two approaches. The systematic review from~}\cite{kampenes2007systematic}\revised{ confirms that the comparison of the median values and the use of box plots are non-parametric measures used frequently for the analysis of software engineering experiments.}
\end{enumerate}
\revised{In line with the established practice, we used the Shapiro-Wilk test and the three non-parametric methods with the following thresholds: a Shapiro-Wilk test $p$ value below $0.05$ as indicating non-normality; and a Wilcoxon signed-rank test $p$ value below $0.05$, a probability of superiority above $0.5$ and a negative median difference to indicate that VERACITY requires a lower testing budget than the baseline approach.}

To assess the generality of \acronym, we report experimental results from two case studies in which \acronym\ was applied to software systems from different domains. The first case study was based on the TAS system from our motivating example. In the second case study, we applied \acronym\ to the verification of an online shopping web application. This system is introduced in Section~\ref{subsect:webapp}, followed by descriptions of the experiments carried out to address the three research questions in Sections~\ref{subsect:RQ1}, \ref{subsect:RQ2} and \ref{subsect:RQ3}, and by a discussion of threats to validity in Section~\ref{subsect:threats}. 

To enable the reproducibility of our results, we have made all the models, properties and data from our experiments available on the \acronym\ project website at \url{https://www.cs.york.ac.uk/tasp/VERACITY}, which also presents a third case study that uses VERACITY for a model with more complex parametric probability distributions\rerevised{, and a fourth case study that applies VERACITY to a larger, 18-parameter model}.

\subsection{Online shopping web application \label{subsect:webapp}}

The system we used for the second case study is an online shopping application adapted from~\cite{filieri2012formal}. 
We modelled the shopping process implemented by this application using a parametric Markov chain that comprises a combination of known and unknown transition probabilities. The known transition probabilities correspond to application components that have been in use for a long time, and for which the values of these probabilities can be determined from application logs. In contrast, the unknown transition probabilities correspond to new versions of several components that the online shopping company's developers have re-implemented and want to evaluate through \emph{A/B testing}.

A/B testing~\cite{fabijan2018experimentation,kohavi2017online,siroker2013b} is a method for testing a new online-application feature, or a new implementation of an existing feature. Frequently used by leading companies like Amazon, Facebook, Google and Microsoft, the method involves splitting the users of a web application into two sets, such that one set of users is given access to a version of the application that includes the new feature (or the new implementation of a feature), while the other set continues to use the standard version of the application. In this way, A/B testing allows companies to evaluate new features and components, and to decide whether they should be included in their online applications or not. \revised{In our case study, we use A/B testing to motivate the scenario where a system has two different versions, such that one is well known in its behaviour and the other is still open to experimentation.}

For our case study, we assume that the engineers want to verify whether the nonfunctional requirements from Table~\ref{table:Online-shopping-requirements} would be satisfied if
four application components were to be replaced with new variants of those components. Furthermore, we assume that in order to limit the disruption of the user experience that may occur if these requirements are in fact violated, the company wants to perform this verification with as little A/B testing of each of the four new component implementations as possible. 

\begin{table*}
	\renewcommand{\arraystretch}{1.2}
	\centering
	\caption{Nonfunctional requirements for the online shopping application}
	
    \vspace*{1.5mm}
    \begin{sffamily}
    \def\tabcolsep{3.75pt}
	\begin{footnotesize}
		\begin{tabular}{p{0.4cm} p{12.8cm} p{2.8cm}} %
			\toprule
			\textbf{ID} 
			&\textbf{Requirement} & \textbf{PCTL formula}
			\\
			\midrule
			R1
			& The probability that customers complete the shopping process successfully shall be above $\mathit{bound}_1$. & $\mathcal{P}_{>\mathit{bound}_1}[\mathrm{F}\;\mathit{success}]$\\
			R2
			& The probability that the authentication component fails shall be below $\mathit{bound}_2$. & $\mathcal{P}_{<\mathit{bound}_2}[\mathrm{F}\; \mathit{authFail}]$\\
			R3
			& The average number of uses of new components per shopping session$^\dagger$ shall exceed $\mathit{bound}_3$.
			& $\mathcal{R}_{>\mathit{bound}_3}[\mathrm{F}\;\mathit{done}]$\\
			\bottomrule
			\multicolumn{3}{l}{$^\dagger$This is a measure of how far the application users progress with the shopping process before giving up or encountering a component failure.}
		\end{tabular}
	\end{footnotesize}
    \end{sffamily}
	\label{table:Online-shopping-requirements}	
	
	\vspace*{-1.5mm}
\end{table*}

\begin{figure}
    \centering 
    \includegraphics[width=\hsize]{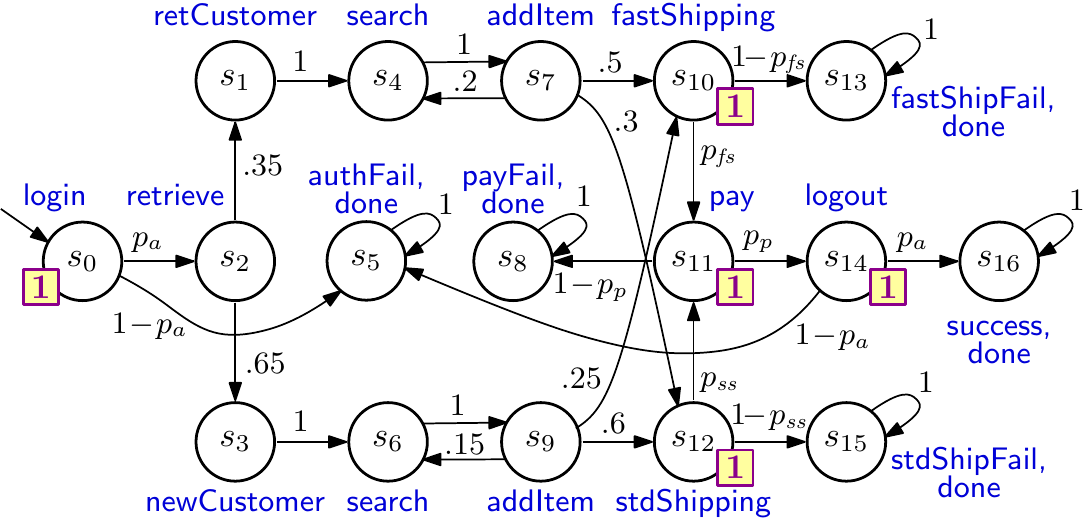}
    
    \vspace*{-1mm}
    \caption{Parametric Markov chain modelling the online shopping application. To enable the verification of requirement \textsf{R3} from Table~\ref{table:Online-shopping-requirements}, the model is augmented with a reward structure that ``counts'' the uses of new components; this reward structure associates a reward $\rho(s_0)=\rho(s_{10})=\rho(s_{11})=\rho(s_{12})=\rho(s_{14})=1$ with each state associated with the use of a new component (as shown in small squares next to these states) and zero rewards with the other states and with all state transitions of the Markov chain.  \label{fig:OnlineShopping-PMC}}
    
    \vspace*{-1.5mm}
\end{figure}

The parametric Markov chain modelling the operation of the online shopping application is shown in Figure~\ref{fig:OnlineShopping-PMC}.
In the initial state ($s_0$)  a customer  
attempts to login. We assume that the authentication web page is one of the components for which a new implementation needs to be tested, and therefore the probability that the customer can follow the authentication instructions and succeeds to login, denoted $p_\mathit{a}$, is unknown. If the authentication succeeds (state $s_2$), the customer is identified either as a returning customer (whose settings from the previous shopping session are restored, state $s_1$) or as a new customer (for whom default settings are used, state $s_3$). In both cases, the customer searches for items to purchase (states $s_4$/$s_6$) and adds them to the shopping basket (states $s_7$/$s_9$), until eventually all the required items are in the shopping basket and the customer moves to checkout where they select between two shipping options: fast shipping (state $s_{10}$) or standard shipping (state $s_{12}$). We assume that the probabilities of the incoming transitions into states $s_1$, $s_3$, $s_4$,  $s_6$, $s_7$, $s_9$, $s_{10}$ and $s_{12}$ are known (from the previous use of the web application) and have the values from Figure~\ref{fig:OnlineShopping-PMC}.

However, we assume that the web application components for selecting the two shipping options have been re-implemented, and therefore the probabilities $p_\mathit{fs}$ and $p_\mathit{ss}$ that the customer manages to use them successfully and to reach the payment state $s_{11}$ are unknown. Likewise, we consider that a new version of the payment component has been implemented, and that the probability $p_p$ that the customer manages to use it successfully (and to move to the logout state $s_{14}$) is unknown. Finally, we assume that the logout involves the use of the same new authentication component that was used for login, and therefore its (unknown) probability of succeeding is $p_a$. 

\subsection{Research question RQ1 \label{subsect:RQ1}}

For each of the two systems used in our evaluation, we synthesised and examined a broad range of simulated scenarios (assuming unit component testing costs $\mathit{cost}_1=\mathit{cost}_2=\ldots=\mathit{cost}_m=1$). \revised{In these scenarios, each parameter} of the Markov chain $M$ from Figure~\ref{fig:approach} \revised{was given a deterministic value (drawn randomly from the interval $[0,1]$) and test outcomes were sampled accordingly. Likewise,} the bounds from the nonfunctional requirements~\eqref{eq:reqs} \revised{were given fixed values drawn randomly from the interval $[0,1]$.} \revised{We synthesised a sufficiently large number of scenarios to} ensure that \revised{the impact of stochasticity in the experiments is low, and that} the evaluation covered a combination of:
\begin{enumerate}
    \item scenarios in which all requirements were satisfied: (i)~by a narrow margin, i.e., the actual values of the properties from the nonfunctional requirements~\eqref{eq:reqs} were close to their associated bounds; (ii)~by a wide margin; and (iii)~some by a narrow margin and the others by a wide margin;
    \item scenarios in which some of the requirements were satisfied and the remaining requirements were violated by: (i)~a narrow margin; (ii)~a wide margin; and (iii)~some by a narrow margin and the others by a wide margin;
    \item scenarios in which all requirements were violated by: (i)~a narrow margin; (ii)~a wide margin; and (iii)~some by a narrow margin and the others by a wide margin.
\end{enumerate}

In all the experiments, we used the \acronym\ tool in conjunction with a simulated component-testing script with the characteristics described in Section~\ref{sect:implementation}. This script emulated the outcome of unit testing the SUV components by using a separate Java pseudorandom number generator for each component. To avoid any bias in the comparison of \acronym\ with the baseline approach mentioned in research question RQ1, we used the same pseudorandom number generator seeds in the corresponding experiments for the two approaches.

\paragraph{Case Study~1 (TAS)}

For the TAS system, we carried out experiments that examined the effectiveness of \acronym\ for a set of 33~scenarios that were randomly generated as described at the beginning of this section. For each of these scenarios, the verification of the TAS nonfunctional requirements was carried out at three confidence levels: $\alpha=0.90$, $\alpha=0.95$ and $\alpha=0.99$. Finally, to answer research question RQ1, two experiments were performed for each scenario and each confidence level~$\alpha$: one in which we used the \acronym\ uncertainty reduction heuristic, and one in which we used the baseline approach that partitions the testing budget of each verification round equally among the TAS components. In total, we performed 198~verification experiments, corresponding to 33~scenarios $\times$ 3~confidence levels $\times$ 2~uncertainty reduction methods.

\begin{figure*}
\centering
    \hspace*{-3mm}
    \mbox{
    \includegraphics[width=0.3\hsize]{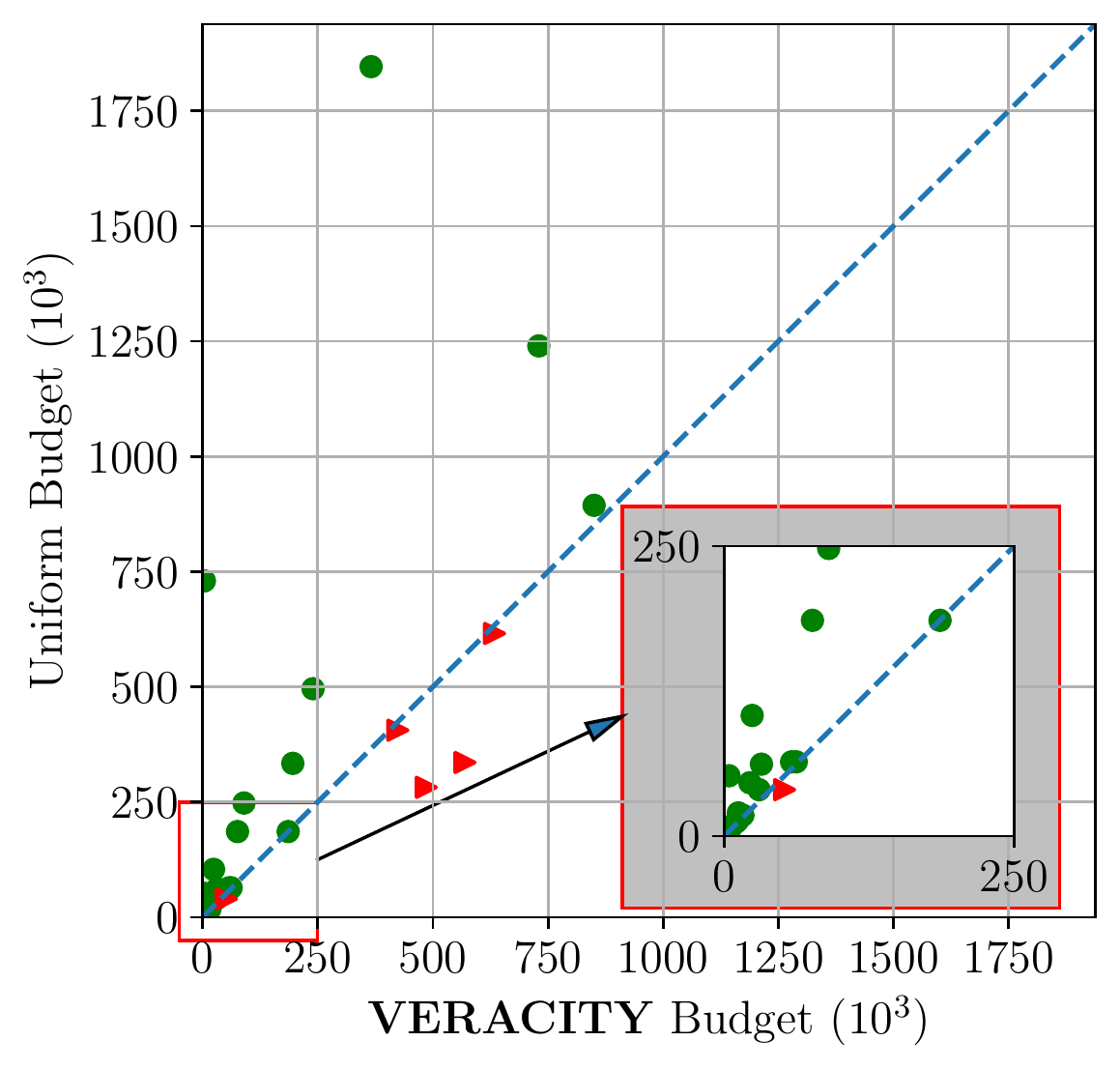}
    \includegraphics[width=0.3\hsize]{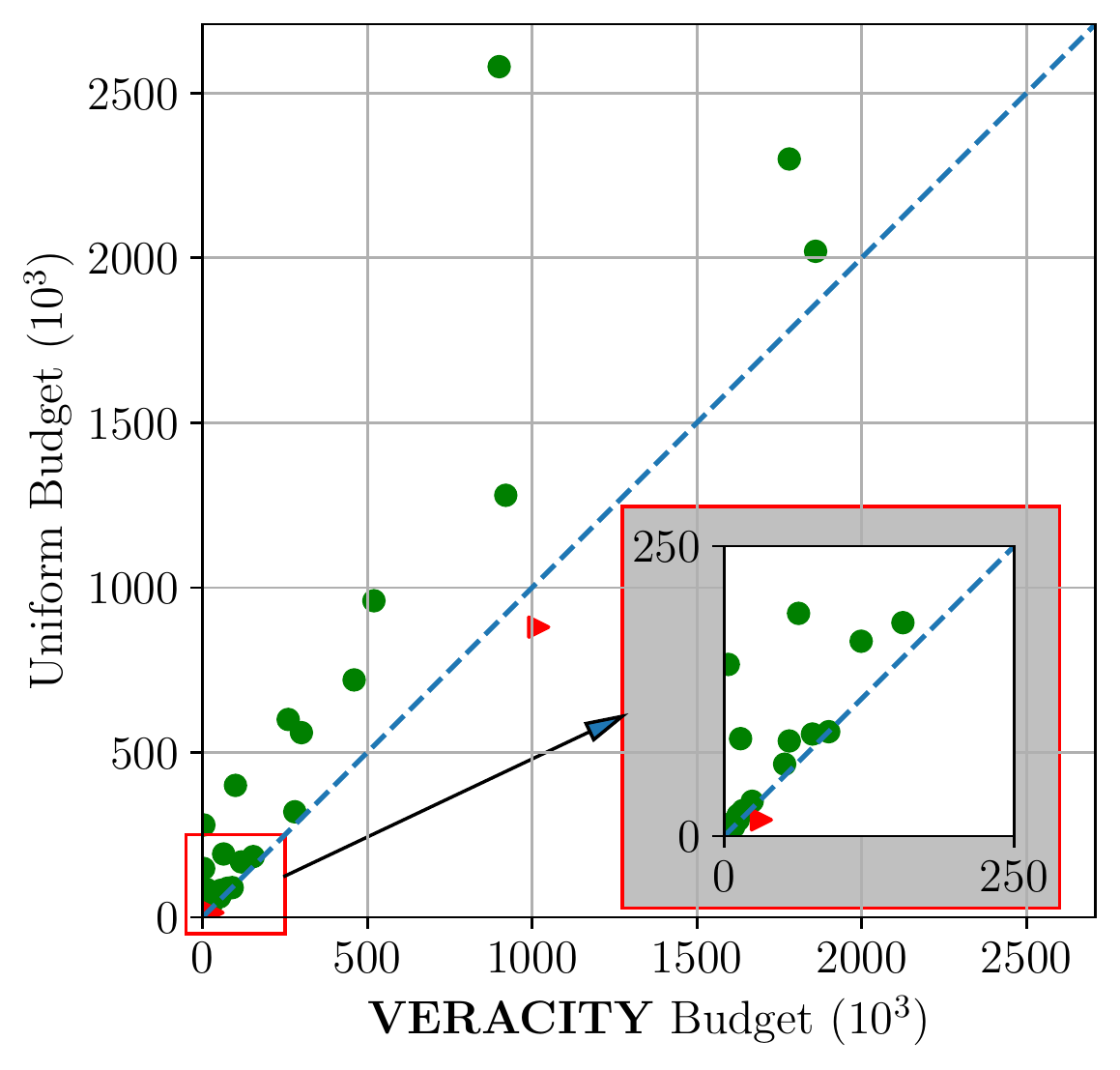}
    \includegraphics[width=0.3\hsize]{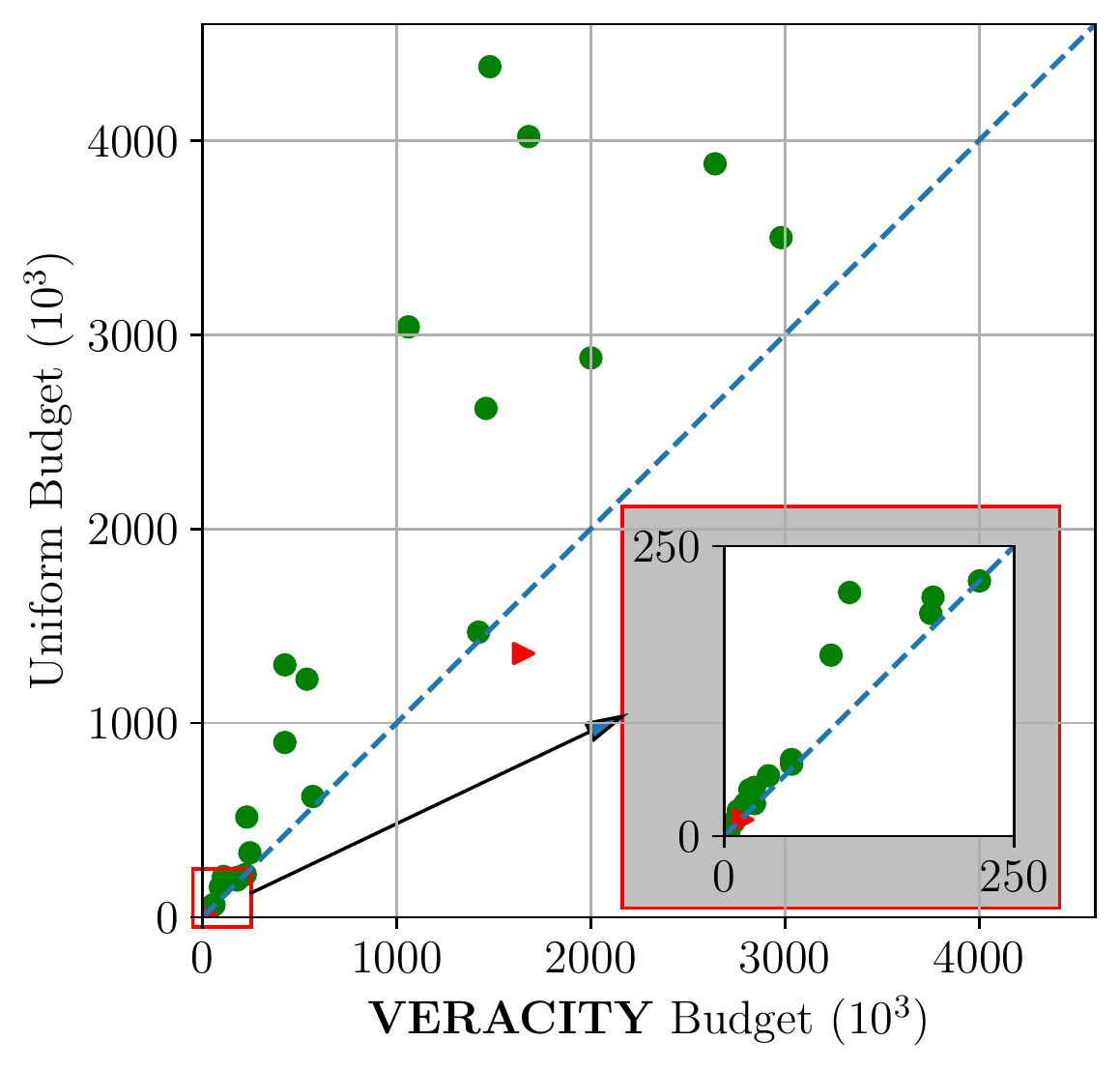}
    }
    
    \begin{footnotesize}
    \hspace*{0\hsize}(a)~$\alpha=0.90$ \hspace*{0.215\hsize}(b)~$\alpha=0.95$  \hspace*{0.215\hsize}(c)~$\alpha=0.99$ 
    \end{footnotesize}
    \caption{\revised{[RQ1:TAS]} Testing budgets required to complete the verification of the TAS nonfunctional requirements using the \acronym\ and the uniform methods for partitioning the testing round budget among the components of the TAS system. The wide range of budgets required to complete the verification process for different scenarios reflects the variety of these scenarios: in some scenarios, the TAS requirements are satisfied or violated by a wide margin (so less testing is needed, as shown by the inset diagrams), whereas in others some or all of the requirements are satisfied or violated by a narrow margin (so much more testing is needed).
    \label{fig:TAS-same-cost}}
\end{figure*}

\begin{figure}
\centering
    \includegraphics[trim=0mm 110mm 10mm 110mm, clip, width=\hsize]{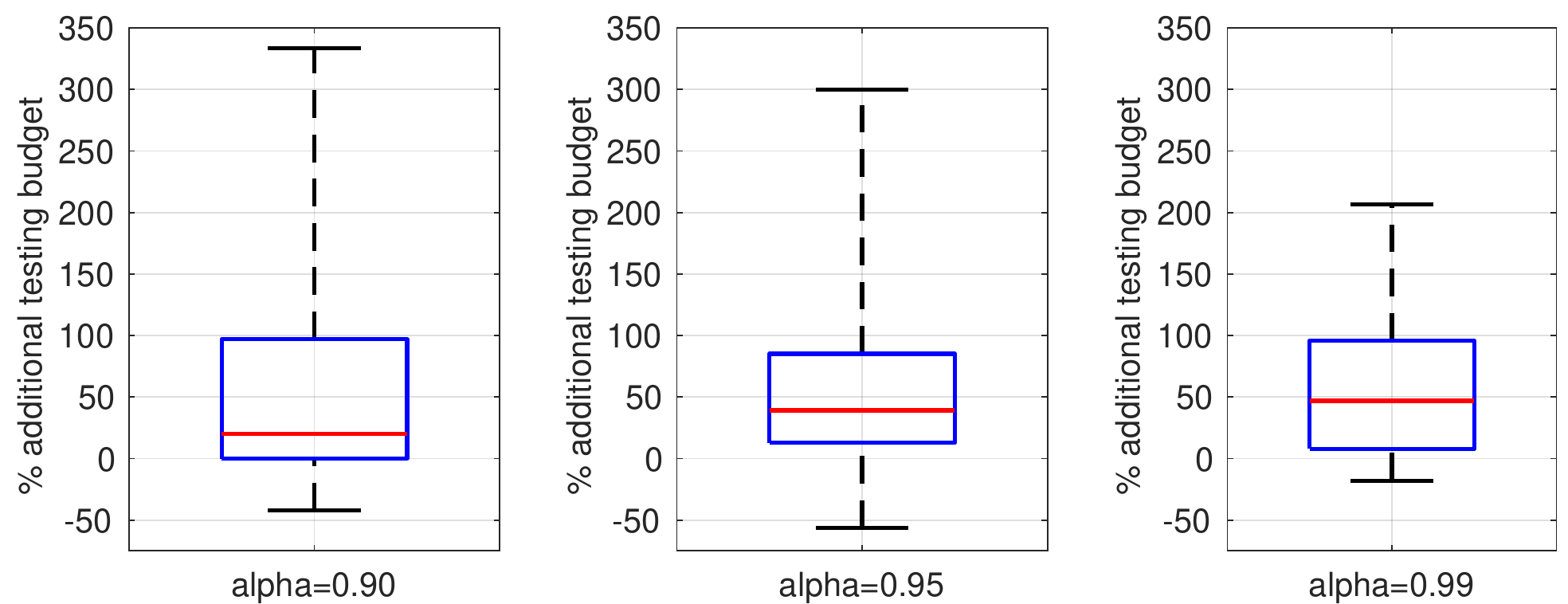}
    \caption{\revised{[RQ1:TAS]} Additional testing budget required to complete the verification of the TAS nonfunctional requirements when the round budget is partitioned using the uniform method instead of the \acronym\ method. To ensure readability, the upper part of the boxplots is truncated, meaning that the outliers at 404\%, 1200\% and 18150\% (for $\alpha=0.90$), and at 4252\% and 6900\% (for $\alpha=0.95$) are not shown. No outliers exist below the bottom whisker of any of the boxplots.
    \label{fig:TAS-same-cost-boxplots}}
\end{figure}

\revised{The~ Shapiro-Wilk normality test showed that the 33~experimental data points are not normally distributed for any of the six combinations of uncertainty reduction method and confidence level ($p = 1.73\!\times\! 10^{-7}$,  $p=1.91\!\times\! 10^{-7}$ and $p=2.31\!\times\! 10^{-6}$ for VERACITY with $\alpha=0.90$, $\alpha=0.95$ and $\alpha=0.99$, respectively; and $p = 5.09\!\times\! 10^{-6}$, $p=4.89\times$\\
$10^{-8}$ and $p=3.03\!\times\! 10^{-6}$ for the baseline method with $\alpha=0.90$, $\alpha=0.95$ and $\alpha=0.99$, respectively). Therefore, we compared the VERACITY and baseline outcomes by using the three non-parametric methods mentioned at the beginning of Section~}\ref{sect:evaluation}\revised{, obtaining the following results, all of which confirm the superiority of VERACITY over the baseline approach:}
\begin{enumerate}
    \item \revised{for $\alpha=0.90$, Wilcoxon  $p=0.003$, probability of superiority $0.697$, and median difference $-1988$;}
    \item \revised{for $\alpha=0.95$, Wilcoxon   $p=0.000$, probability of superiority $0.788$, and median difference $-25957$;}
    \item \revised{for $\alpha=0.99$, Wilcoxon $p=0.000$, probability of superiority $0.848$, and median difference $-25902$.}
\end{enumerate}
\revised{As mentioned at the beginning of Section~}\ref{sect:evaluation}\revised{, we also provide scatter plots and box plots enabling the visual inspection of the experimental data} (Figures~\ref{fig:TAS-same-cost} and~\ref{fig:TAS-same-cost-boxplots}).

These results show that \acronym\ outperforms the baseline method by completing the verification process with smaller testing budgets for a great majority of the scenarios and at all confidence levels. In a few scenarios, the baseline method performs better than \acronym\ (typically only marginally better). This is expected given the stochastic nature of the verified system, and the fact that the verification starts with no knowledge about the behaviour of the three TAS components. Finally, in a small number of additional scenarios, \acronym\ achieves only modest testing cost savings. This is also expected, as the best way to reduce epistemic uncertainty in some verification scenarios is to partition the round testing budget approximately equally among the tested system components, and our approach manages to do this well.

The experimental results show that the testing budget reductions enabled by \acronym\ are particularly significant when the verification is carried out at higher confidence levels. This is extremely useful for two reasons. First, in real-world scenarios, the nonfunctional requirements of software systems should be verified with high levels of confidence ($\alpha=0.99$ \revised{or even higher, as in e.g. medicine}); deploying a system whose requirements were only verified at a low confidence level introduces significant risks. Second, the testing budget needed to complete the verification increases with the confidence level $\alpha$, as the epistemic uncertainty needs to be reduced much more in order to make decisions with higher confidence. This increase of the required testing budget for larger  $\alpha$ values is clearly visible in the scales of the graphs from Figure~\ref{fig:TAS-same-cost}. As such, the scenarios in which \acronym\ reduces the cost of testing the most are: (i)~of particular practical importance; and (ii)~characterised by high testing costs, so the cost reductions achieved by our uncertainty reduction method are especially beneficial.

\paragraph{Case Study~2 (Online shopping web application)}

To assess the effectiveness of \acronym\ for the online shopping web application (WebApp), we performed a similar suite of experiments to those described for the TAS case study. This time, we examined the ability of \acronym\ to reduce testing costs compared to the baseline uncertainty reduction method for the verification of 30~randomly generated scenarios. In each of the 30~scenarios, the verification of the WebApp requirements was carried out at three confidence levels ($\alpha=0.90$, $\alpha=0.95$ and $\alpha=0.99$), for both the \acronym\ and the uniform uncertainty reduction methods, giving a total of 180~experiments.

\begin{figure*}
\centering
    \hspace*{-3mm}
    \mbox{
    \includegraphics[width=0.3\hsize]{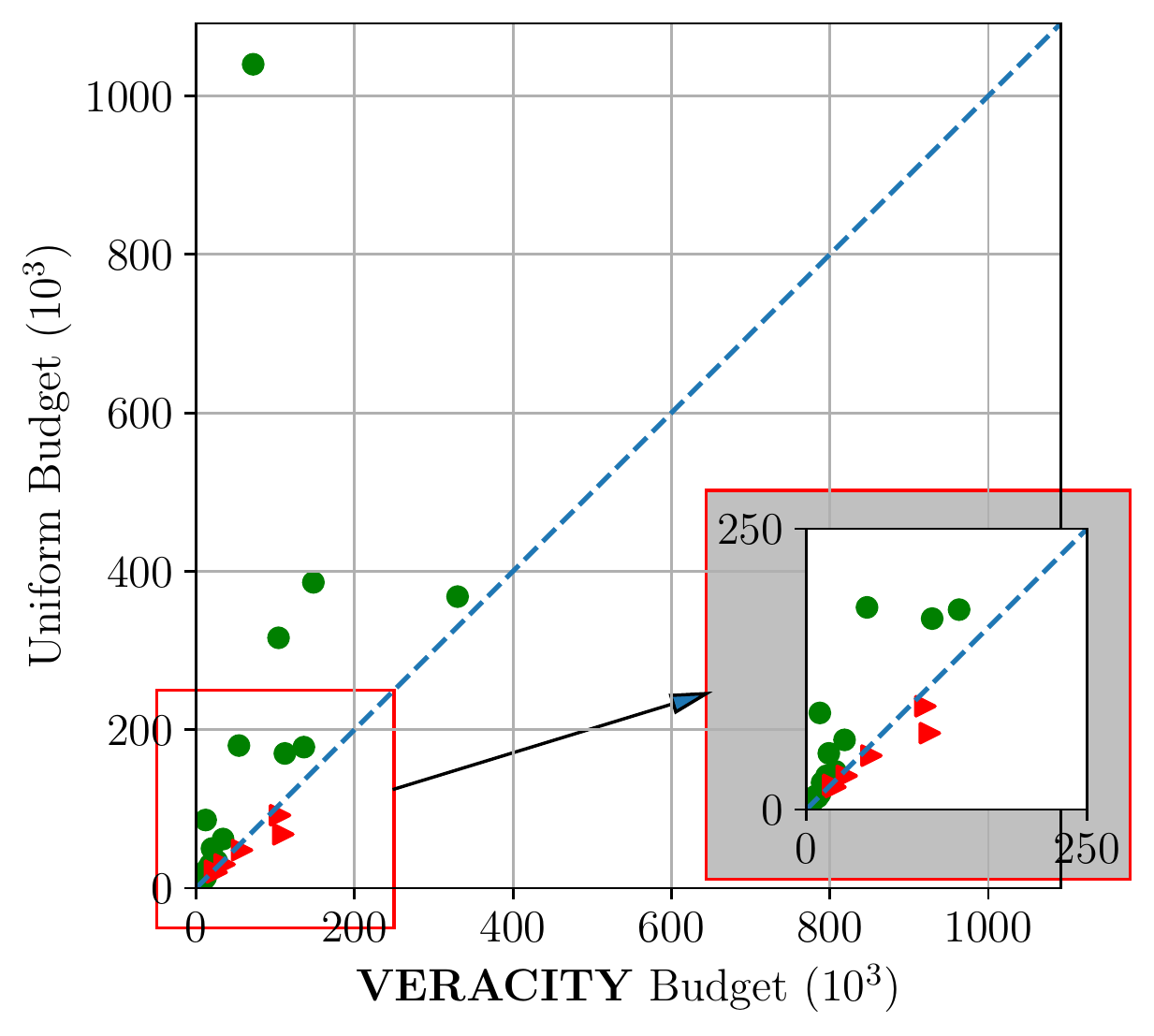}
    \includegraphics[width=0.3\hsize]{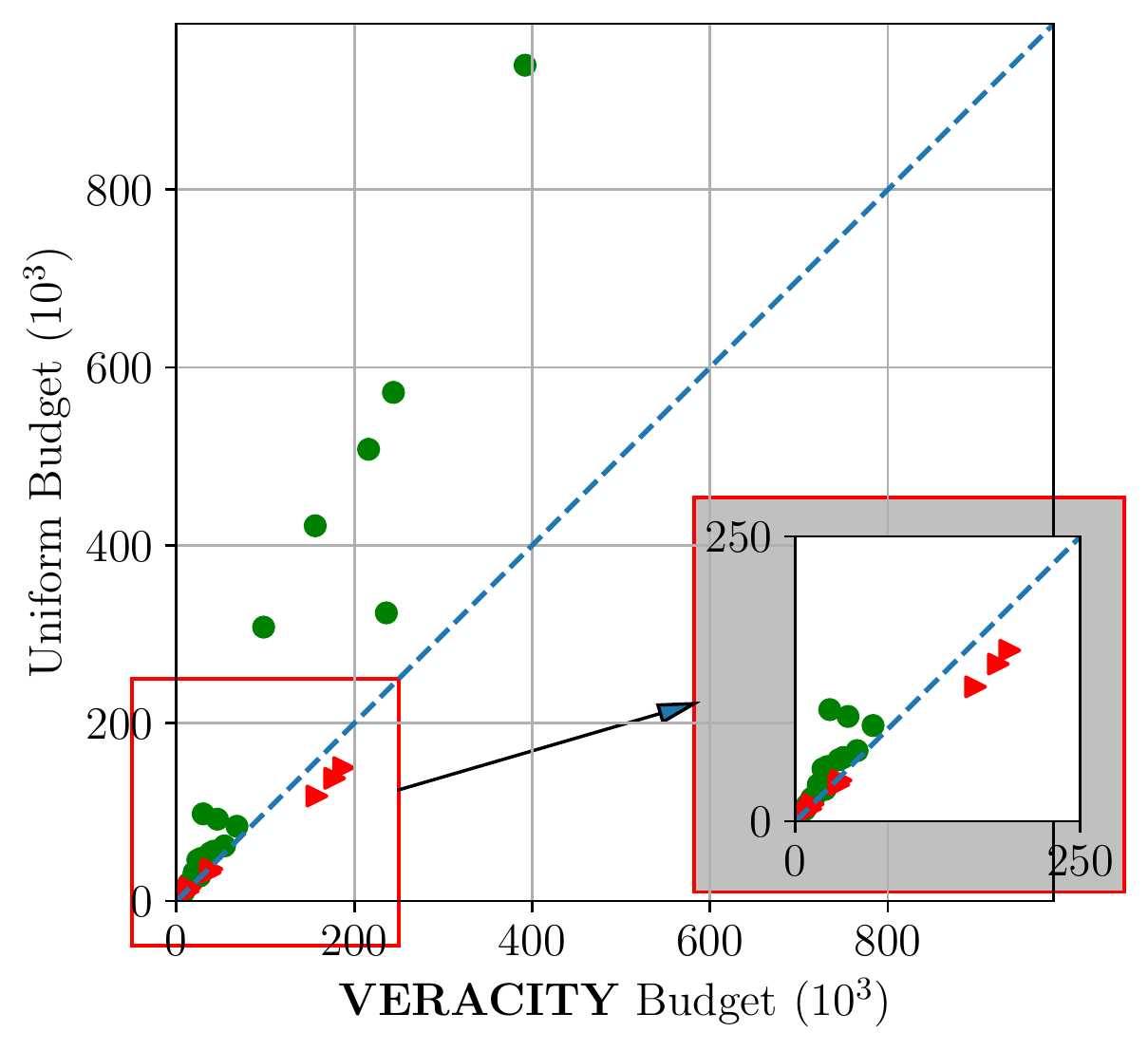}
    \includegraphics[width=0.3\hsize]{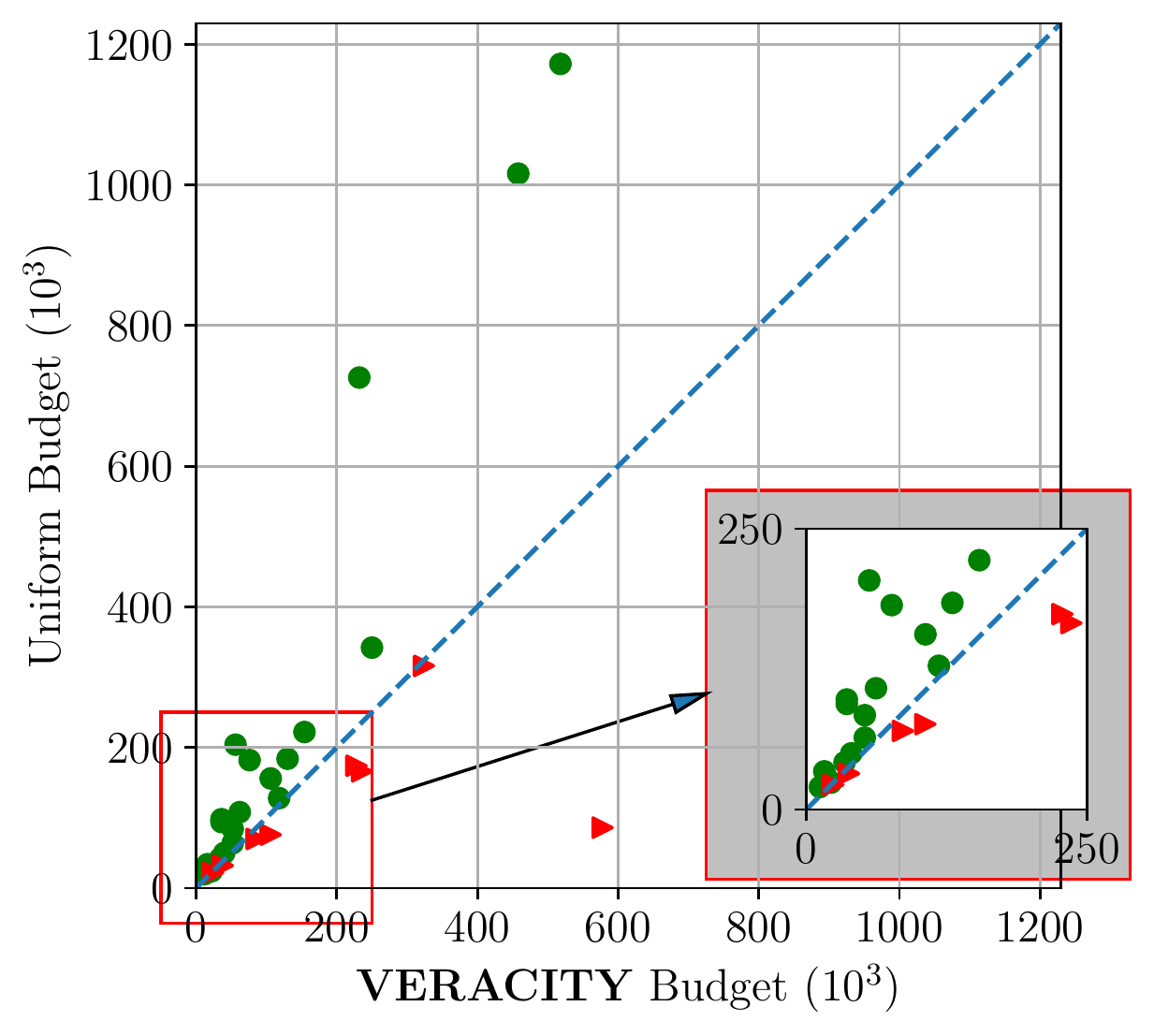}
    }
    
    \begin{footnotesize}
    \hspace*{0\hsize}(a)~$\alpha=0.90$ \hspace*{0.215\hsize}(b)~$\alpha=0.95$  \hspace*{0.215\hsize}(c)~$\alpha=0.99$ 
    \end{footnotesize}
    \caption{\revised{[RQ1:WebApp]} Testing budgets required to complete the verification of the WebApp nonfunctional requirements using the \acronym\ and the uniform methods for partitioning the testing round budget among the components of the online shopping system.  \label{fig:web-app-same-cost}}
\end{figure*}

\revised{The 30~experimental data points are not normally distributed for any combination of uncertainty reduction method and confidence level (Shapiro-Wilk test $p=6.35\times 10^{-7}$, $p=9.30\times 10^{-6}$, and $p=1.44\times 10^{-6}$  for VERACITY with $\alpha=0.90$, $\alpha=0.95$ and $\alpha=0.99$, respectively; and $p=1.30$ $\times 10^{-8}$, $p=3.41\times 10^{-7}$, and $p=2.33\times 10^{-7}$ for the baseline method with $\alpha=0.90$, $\alpha=0.95$ and $\alpha=0.99$, respectively). The Wilcox signed-rank test, the probability of superiority measure, and the median difference all indicate that VERACITY outperforms the baseline method for every confidence level:}
\begin{itemize}
    \item \revised{for $\alpha=0.90$, Wilcoxon  $p=0.001$, probability of superiority $0.800$, and median difference $-5000$;}
    \item \revised{for $\alpha=0.95$, Wilcoxon  $p=0.002$, probability of superiority $0.733$, and median difference $-10000$;}
    \item \revised{for $\alpha=0.99$, Wilcoxon  $p=0.018$, probability of superiority $0.700$, and median difference $-11000$};
\end{itemize}

\begin{figure}
\centering
    \includegraphics[trim=0mm 110mm 10mm 110mm, clip, width=\hsize]{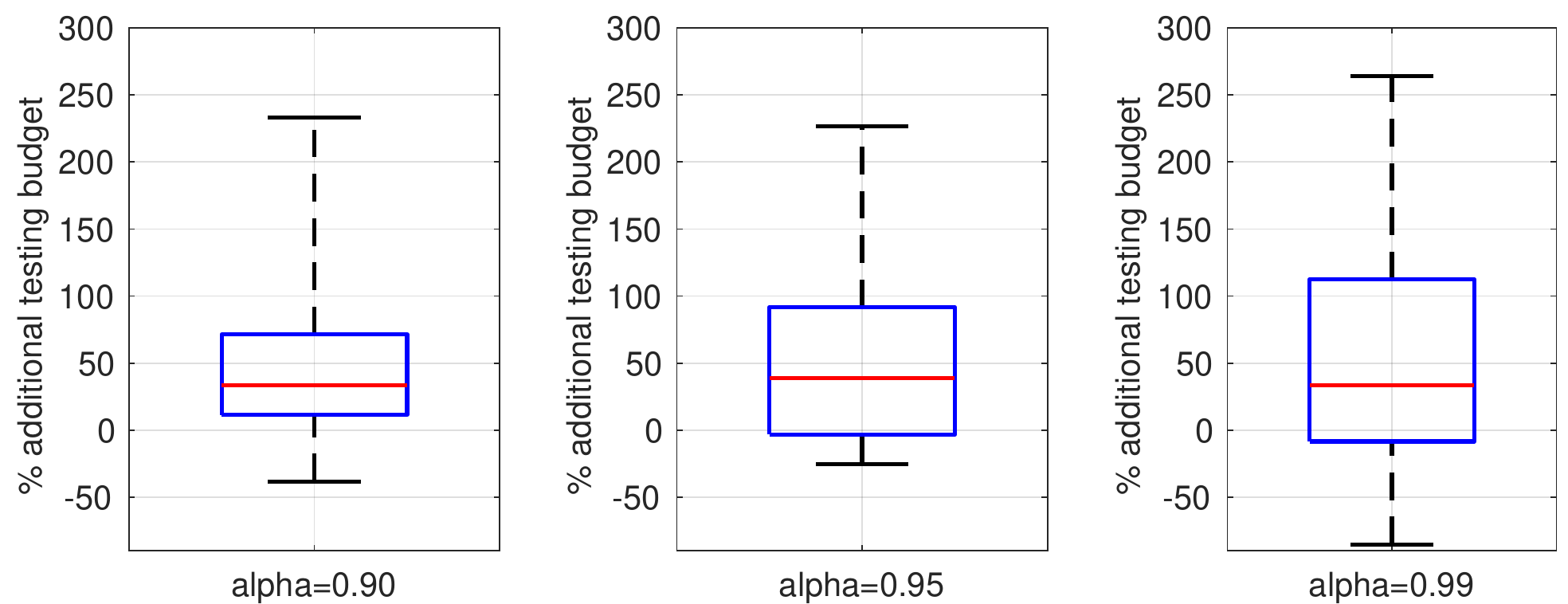}
    
    \vspace*{-1mm}
    \caption{\revised{[RQ1:WebApp]} Additional testing budget required to complete the verification of the WebApp nonfunctional requirements when the round budget is partitioned using the uniform method instead of the \acronym\ method. To ensure readability, the upper part of the boxplots is truncated at 300\%, meaning that the outliers at 616\% and 1344\% (for $\alpha=0.90$) are not shown. No outliers exist below the bottom whisker of any of the boxplots.
    \label{fig:web-app-same-cost-boxplots}}
    
    \vspace*{-2mm}
\end{figure}

Figures~\ref{fig:web-app-same-cost} and~\ref{fig:web-app-same-cost-boxplots} summarise the results of these experiments. As for the TAS system, \acronym\ successfully reduces the testing budget required to complete the verification of the nonfunctional requirements, across a wide range of testing budget needs (where small testing budgets are needed when the requirements are satisfied or violated by a wide margin, and large budgets when some or all of the requirements are narrowly satisfied/violated). In the small number of scenarios where the uniform round budget partitioning method achieves better results, the overall testing budget is small, and the \acronym-based verification is typically only marginally more expensive. Again, many significant budget reductions occur when (i)~the baseline method budget is high and (ii)~the requirements are verified at higher confidence levels. For instance, all of the baseline method budgets above 400,000 from Figure~\ref{fig:web-app-same-cost} (one for $\alpha=0.90$, four for $\alpha=0.95$, and three for $\alpha=0.99$) are at least halved by \acronym.

\paragraph{\revised{Discussion}}

\revised{VERACITY is a heuristic whose behaviour depends on the configuration parameters $\epsilon_1$ and $\epsilon_2$ from Algorithm~}\ref{algorithm:heuristic}\revised{, on the configuration of the formal verification with confidence intervals step from Figure~}\ref{fig:approach}\revised{, and on the stochasticity of its component testing outcomes. As such, it is expected that VERACITY cannot always outperform the baseline method, just like a superior medical treatment is unfortunately not always outperforming an inferior treatment. Therefore, we used established analyses methods from the software engineering domain to evaluate VERACITY, and the results of these analyses show its superiority over the baseline method. Additionally, our examination of scenarios in which VERACITY used larger testing budgets than the baseline method revealed that these were typically scenarios in which:}
\begin{itemize}
    \item \revised{the uniform budget partitioning used by the baseline method was the (nearly) optimal option, so the best VERACITY could have achieved was to match the performance of the baseline method---but this would have required perfectly chosen values for the VERACITY configuration parameters;}
    \item \revised{the stochastic effects of the component testing required for the verification had an adverse effect on VERACITY;}
    \item \revised{calibrating the values of the VERACITY configuration parameters (instead of using their default values) can improve the performance of our method;}
    \item \revised{a combination of the previous three factors was at play.}
\end{itemize}
\revised{The influence of such factors is unavoidable for software engineering methods that employ heuristics, and---given the positive results of the analyses presented earlier in this section and in the rest of our evaluation---it does not affect the usefulness of VERACITY.}

\subsection{Research question RQ2 \label{subsect:RQ2}}

\revised{Experiments similar to those from Section~}\ref{subsect:RQ1}\revised{ (but focusing on varying the component testing costs) were carried out to evaluate the ability of VERACITY to handle the} practical situations \revised{where} the costs of testing different components of a system are different. One situation when this is likely to be true is, for instance, when these costs represent the times required for the regression testing of a software system with several modified components~\cite{MUCCINI20061379}. Another situation when this may apply is for testing different web services at runtime, when a limited overall time (i.e., testing budget) is available to verify whether using these services as part of a service-based system like TAS satisfies a set of nonfunctional requirements~\cite{calinescu2011dynamic}. This is also likely to be true for the A/B testing of new features of an online application~\cite{fabijan2018experimentation,kohavi2017online,siroker2013b} like the shopping application from Section~\ref{subsect:webapp}, where these costs may represent the different (expected) business impact of each of the new features not working as intended.

\begin{figure*}
\centering
    \hspace*{-3mm}
    \mbox{
    \includegraphics[width=0.3\hsize]{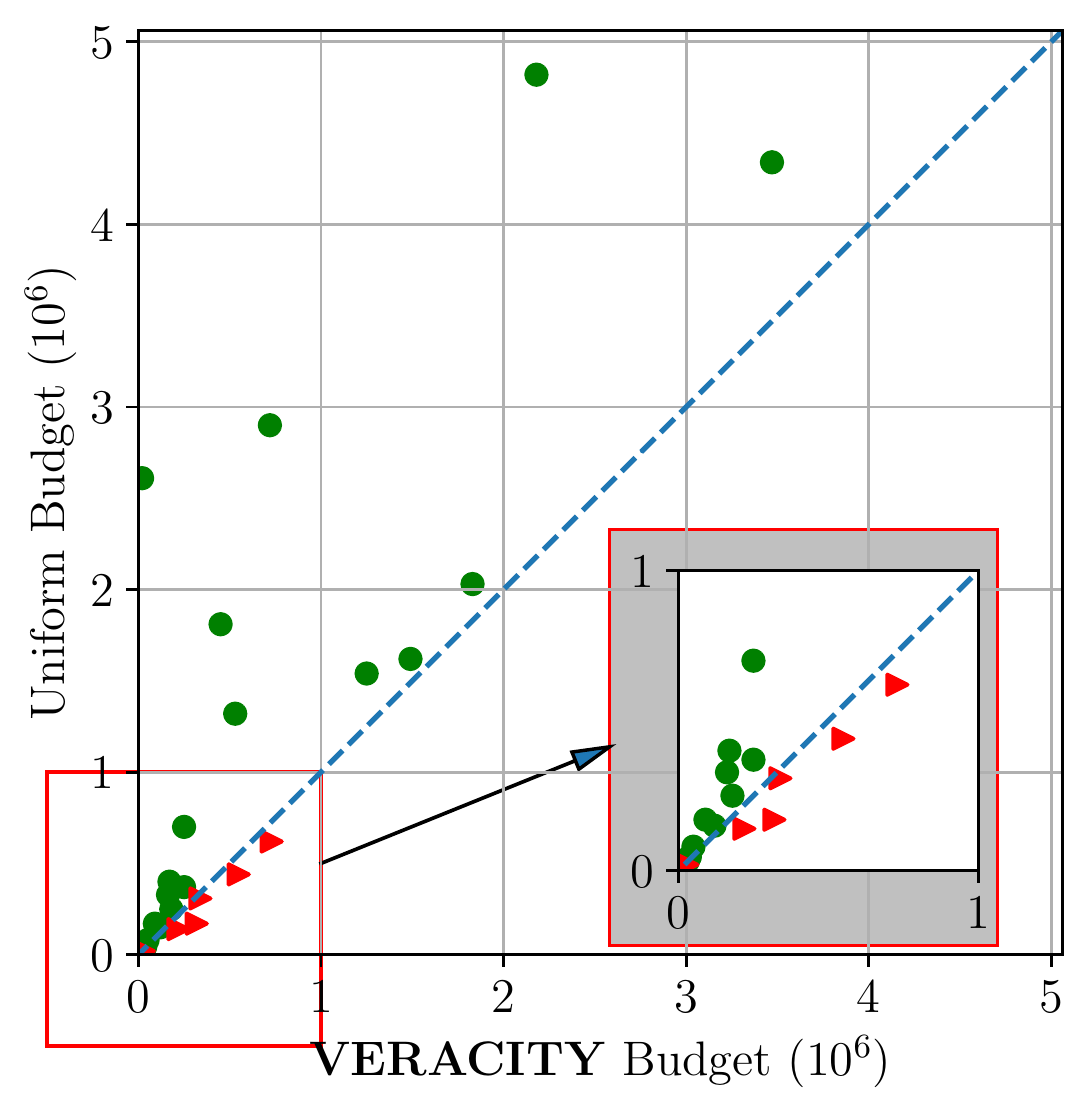}
    \includegraphics[width=0.3\hsize]{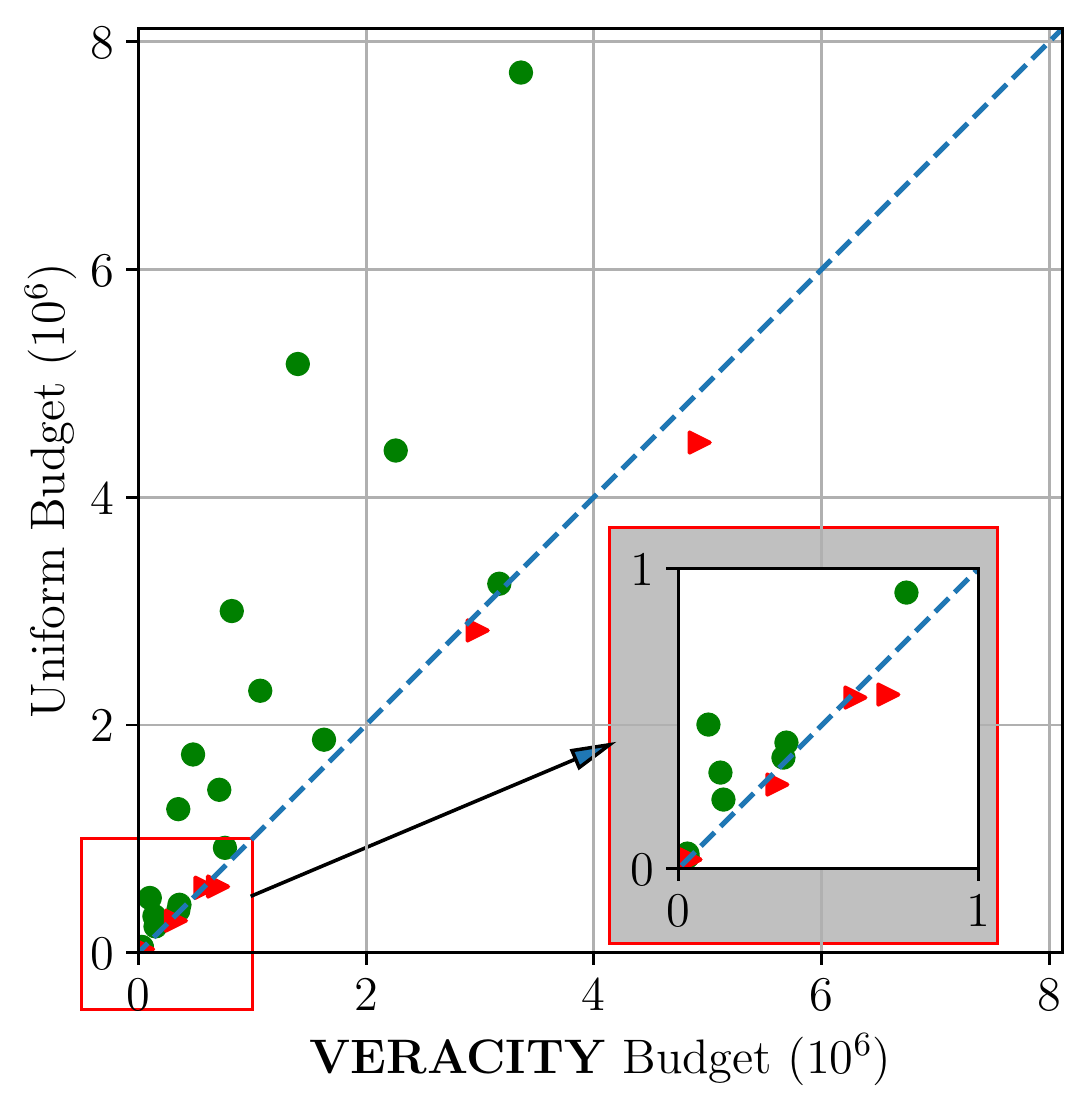}
    \includegraphics[width=0.3\hsize]{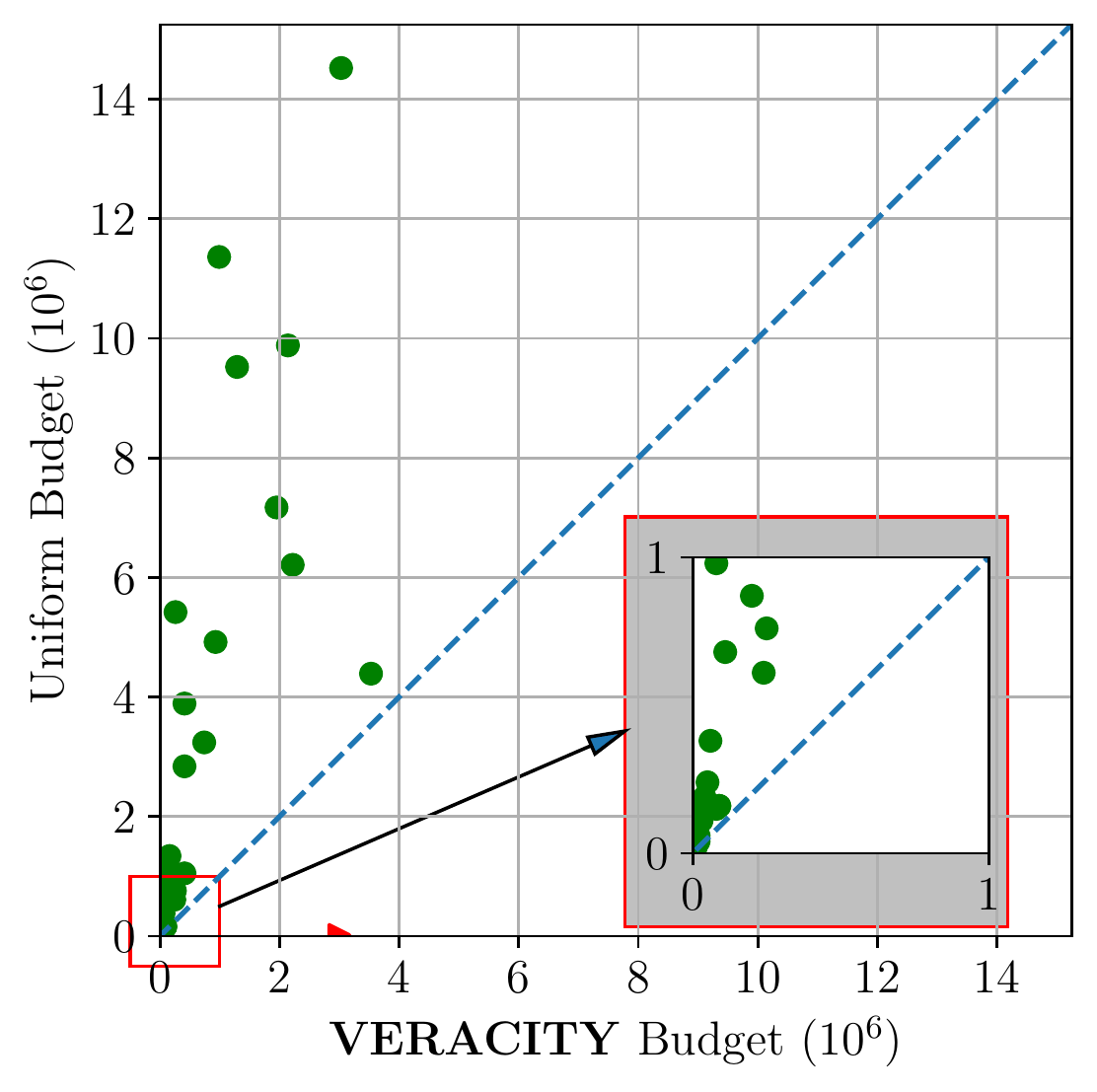}
    }
    
    \begin{footnotesize}
    \hspace*{0\hsize}(a)~TAS, $\alpha=0.90$ \hspace*{0.15\hsize}(b)~TAS, $\alpha=0.95$  \hspace*{0.15\hsize}(c)~TAS, $\alpha=0.99$ 
    \end{footnotesize}
    
    \vspace*{2mm}
    \hspace*{-3mm}
    \mbox{
    \includegraphics[width=0.3\hsize]{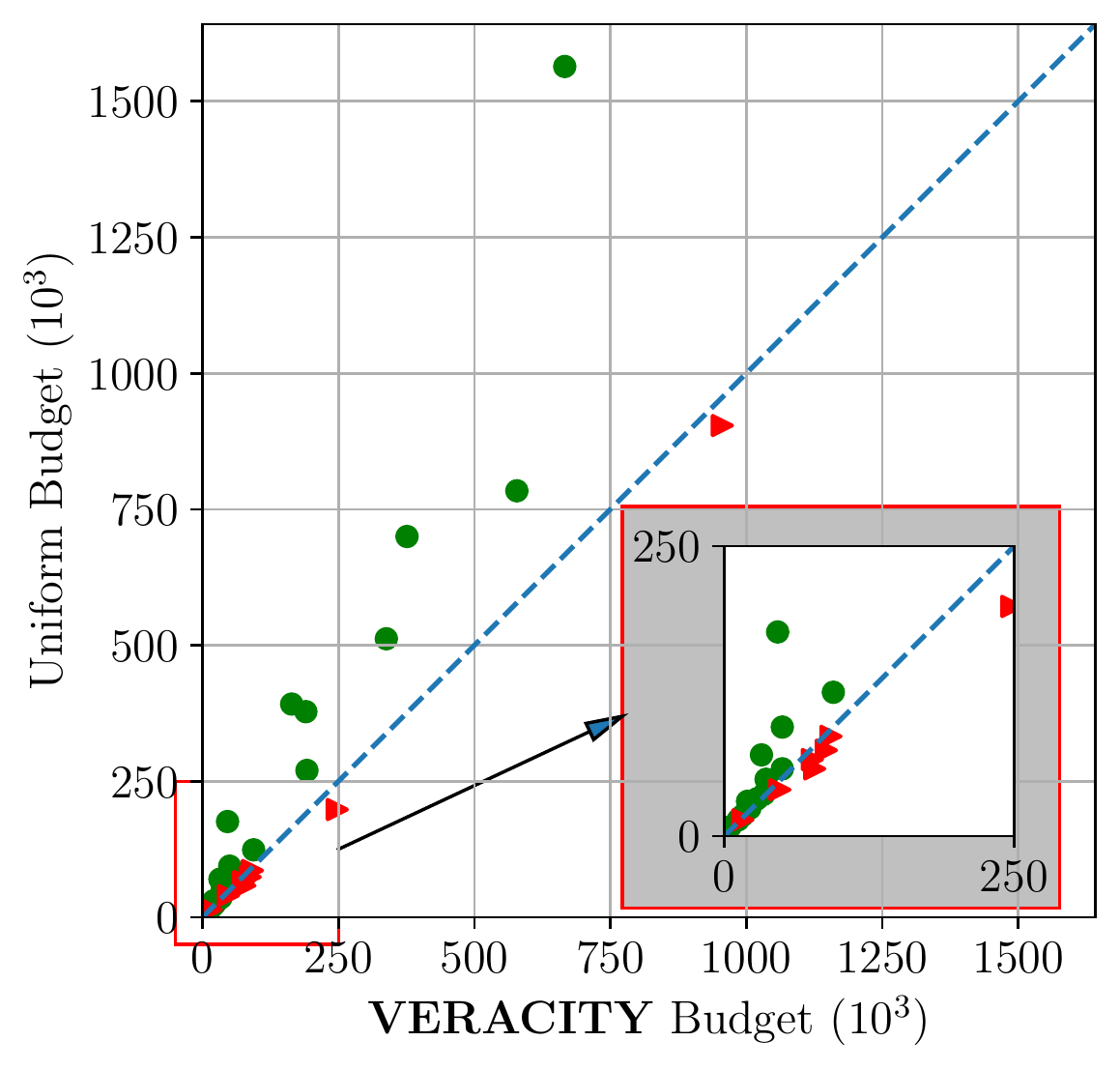}
    \includegraphics[width=0.3\hsize]{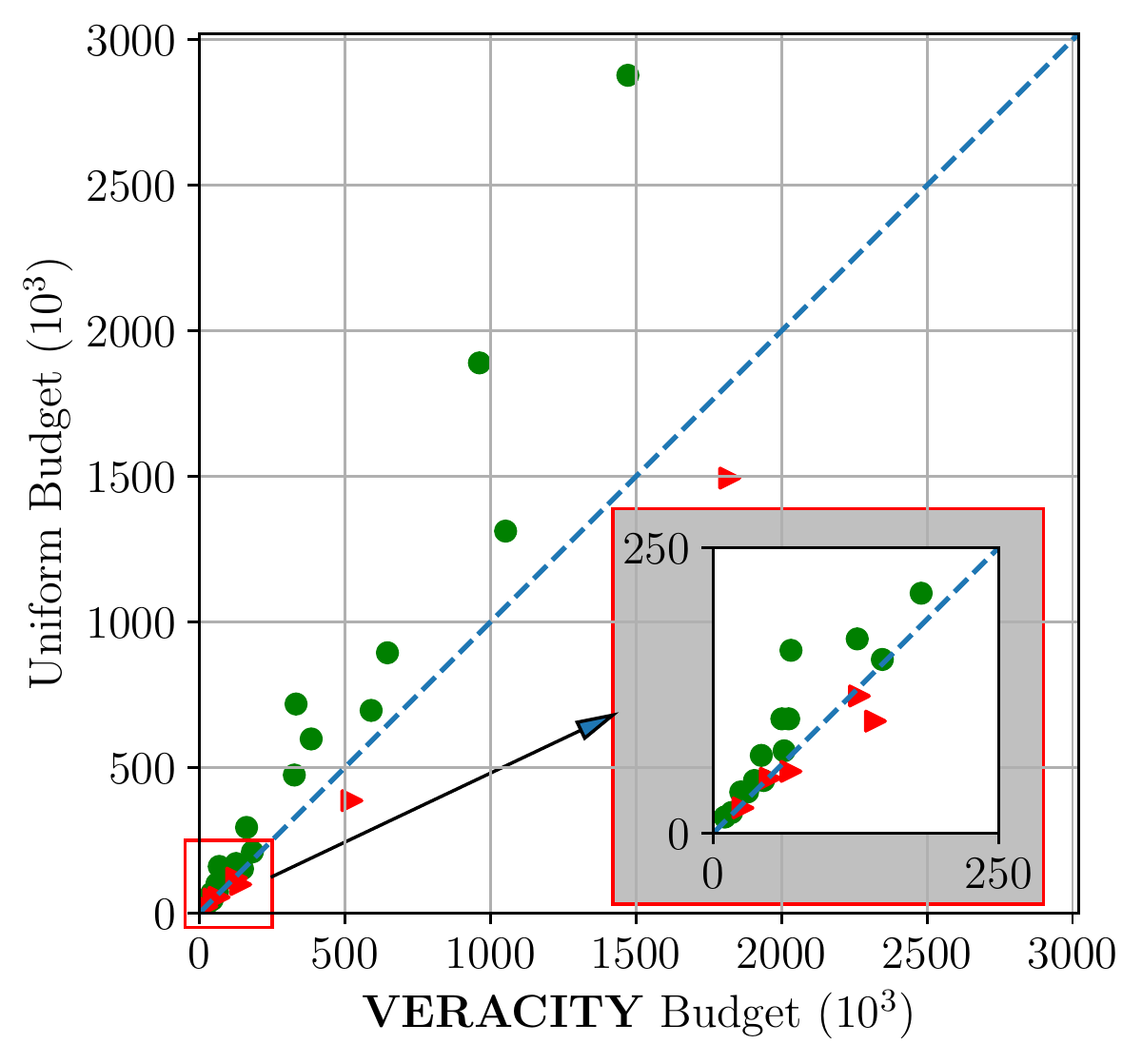}
    \includegraphics[width=0.3\hsize]{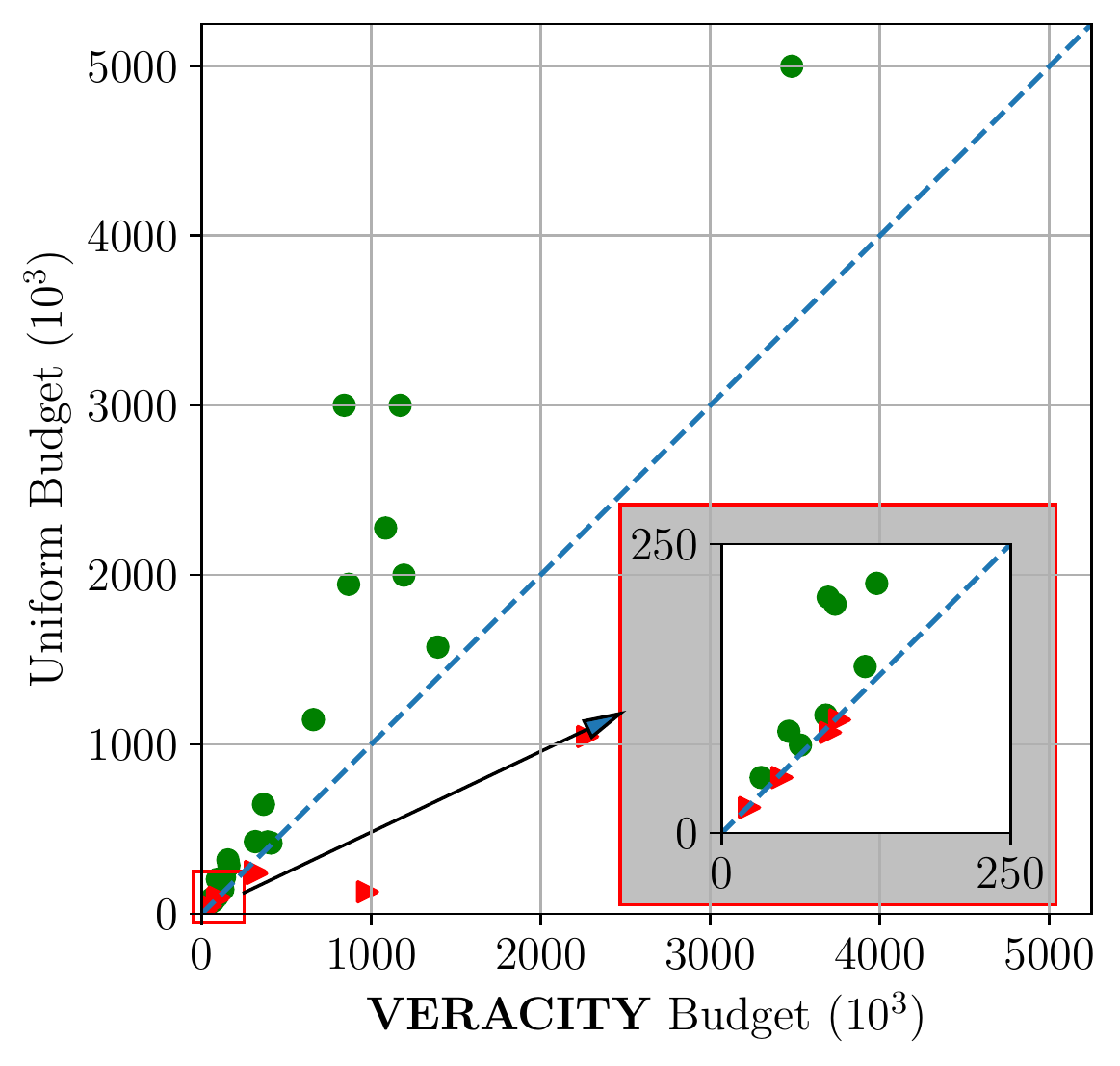}
    }
    
    \begin{footnotesize}
    \hspace*{0\hsize}(d)~WebApp, $\alpha=0.90$ \hspace*{0.1\hsize}(e)~WebApp, $\alpha=0.95$  \hspace*{0.1\hsize}(f)~WebApp, $\alpha=0.99$ 
    \end{footnotesize}
    
    \caption{\revised{[RQ2]} Testing budgets required to complete the verification process using the \acronym\ and the uniform methods for partitioning the testing round budget among the components of the TAS and WebApp systems.  \label{fig:multi-cost}}
    
\end{figure*}

\begin{figure}
\centering
    \includegraphics[trim=0mm 107mm 10mm 110mm, clip, width=\hsize]{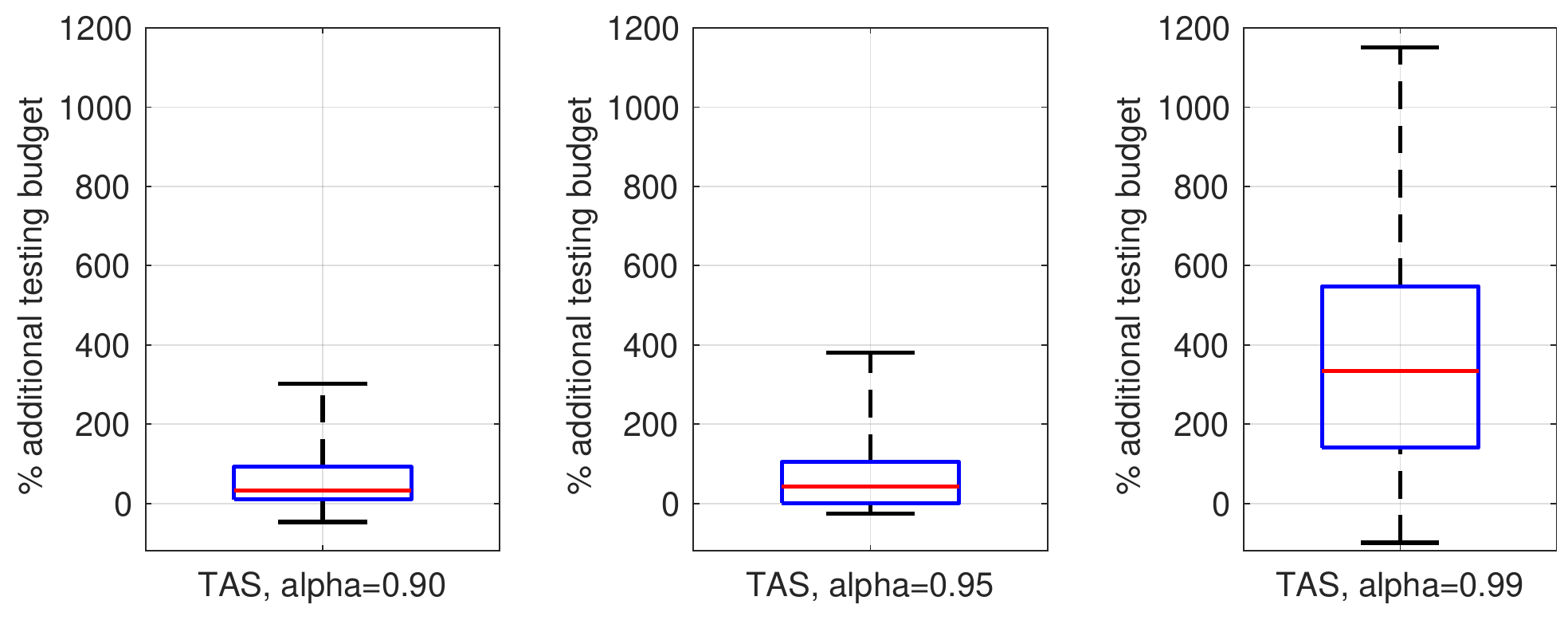}
    
    \includegraphics[trim=0mm 110mm 10mm 110mm, clip, width=\hsize]{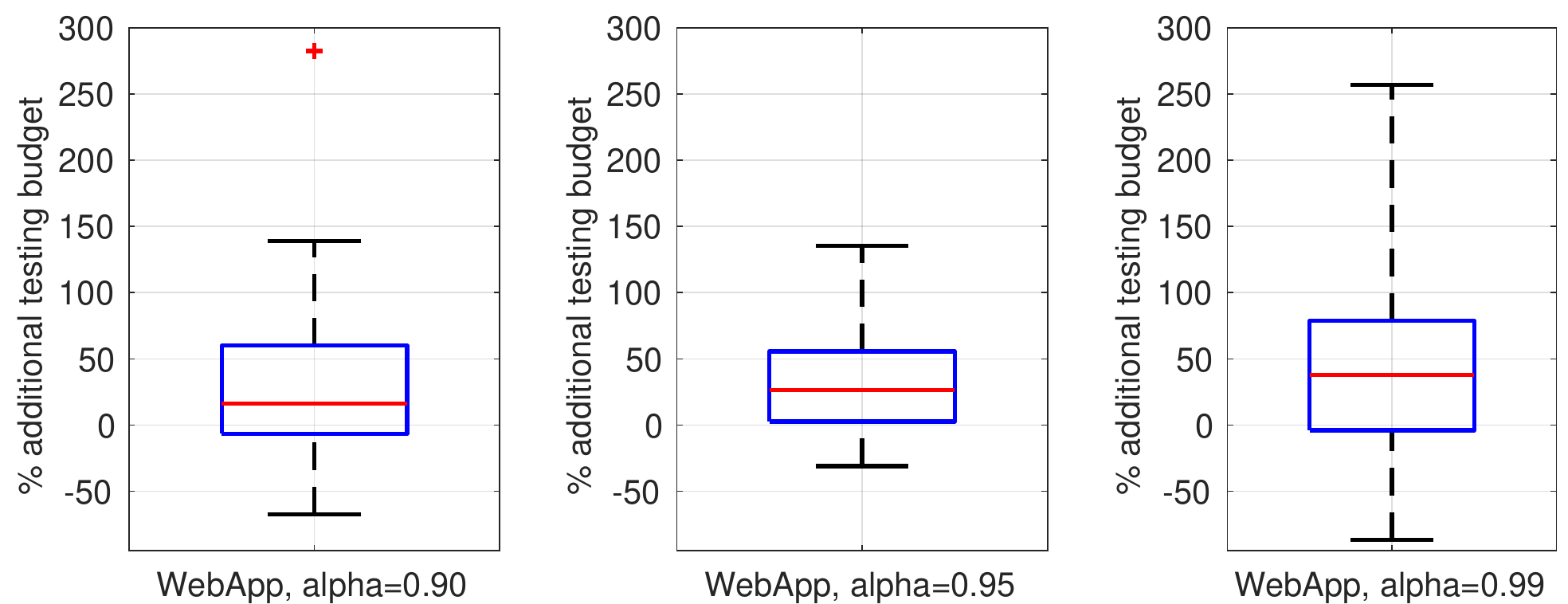}
    
    \caption{\revised{[RQ2]} Additional testing budget required to complete the verification of the TAS and WebApp nonfunctional requirements when the round budget is partitioned using the uniform method instead of the \acronym\ method. To ensure readability, the upper part of the TAS boxplots is truncated at 1200\%, meaning that two TAS outliers at 12950\% (for $\alpha=0.90$) and at 1984\% (for $\alpha=0.99$) are not shown. No other hidden outliers exist for any of the boxplots.
    \label{fig:multi-cost-boxplots}}
    
    \vspace*{-7.5mm}
\end{figure}

To evaluate the usefulness of \acronym\ in such situations, we repeated all the experiments from Section~\ref{subsect:RQ1} assuming different testing costs for the components of the TAS and WebApp systems from our two case studies. To this end, we took each of the 33~TAS verification scenarios and of the 30~WebApp verification scenarios, and we assigned randomly generated testing costs in the interval $[1,5]$ to the three TAS components and the four WebApp components, respectively. 

\revised{The Shapiro-Wilk test indicated that the experimental results were not normally distributed: $p=1.15\times 10^{-7}$, $p=7.64\times 10^{-7}$, and $p=6.58\times 10^{-7}$ for VERACITY applied to TAS at $\alpha=0.90$, $\alpha=0.95$ and $\alpha=0.99$; $p=7.35$ $\times 10^{-7}$, $p=1.23\times 10^{-6}$, and $p=1.99\times10^{-6}$ for VERACITY applied to WebApp at $\alpha=0.90$, $\alpha=0.95$ and $\alpha=0.99$; $p=5.51\times 10^{-7}$, $p=2.97\times 10^{-6}$ and $p=3.98\times 10^{-6}$ for the baseline method applied to TAS at $\alpha=0.90$, $\alpha=0.95$ and $\alpha=0.99$; and $p=6.23\times 10^{-7}$, $p=6.06\times 10^{-7}$ and $p=1.87\times 10^{-6}$ for the baseline method applied to WebApp at $\alpha=0.90$, $\alpha=0.95$ and $\alpha=0.99$. We therefore applied the three non-parametric analysis methods, obtaining the following results for the TAS experiments:}
\begin{itemize}
    \item \revised{for $\alpha=0.90$, Wilcoxon  $p=0.001$, probability of superiority  $0.818$, and median difference $-29973$;}
    \item \revised{for $\alpha=0.95$, Wilcoxon $p=0.002$, probability of superiority $0.758$, and median difference $-19985$;}
    \item \revised{for $\alpha=0.99$, Wilcoxon $p=0.000$, probability of superiority $0.970$, and median difference $-639617$,}
\end{itemize}
\revised{and the results below for the WebApp experiments:}
\begin{itemize}
    \item \revised{for $\alpha=0.90$, Wilcoxon $p$\rerevised{=}$0.024$, probability of superiority $0.700$, and median difference $-4000$;}
    \item \revised{for $\alpha=0.95$, Wilcoxon $p=0.002$, probability of superiority $0.767$, and median difference $-19000$};
    \item \revised{for $\alpha=0.99$, Wilcoxon $p=0.001$, probability of superiority $0.733$, and median difference $-59000$}.
\end{itemize}
\revised{These results indicate the superiority of VERACITY over the baseline approach, at all confidence levels and for both TAS and WebApp.}

The testing budgets required to complete the verification process using the \acronym\ and the baseline round-budget partitioning methods in these scenarios with different component testing costs are compared in Figures~\ref{fig:multi-cost} and~\ref{fig:multi-cost-boxplots}. As in the scenarios with the same testing costs for all components, our \acronym\ verification approach outperforms the baseline verification approach in the majority of the examined scenarios, often by a large margin. As shown in Figure~\ref{fig:multi-cost-boxplots}, this margin increases as $\alpha$ increases. This increase is particularly significant for the TAS system, where the median additional testing budget required by the baseline verification approach grows from 33\% at $\alpha=0.90$ to 42\% at $\alpha=0.95$, and 335\% at $\alpha=0.99$. This growth is less pronounced but still present for the WebApp system, where the median additional testing budget increases from 16\% at $\alpha=0.90$ to 26\% at $\alpha=0.95$, and 38\% at $\alpha=0.99$.

In the small number of scenarios where the baseline approach completes the verification within a smaller testing budget, the difference between this approach and \acronym\ is typically modest, and/or occurs for scenarios where both approaches perform the verification with relatively small overall testing budgets.

\begin{figure*}
\centering
    \includegraphics[trim=0mm 95mm 12mm 100mm, clip, width=0.65\hsize]{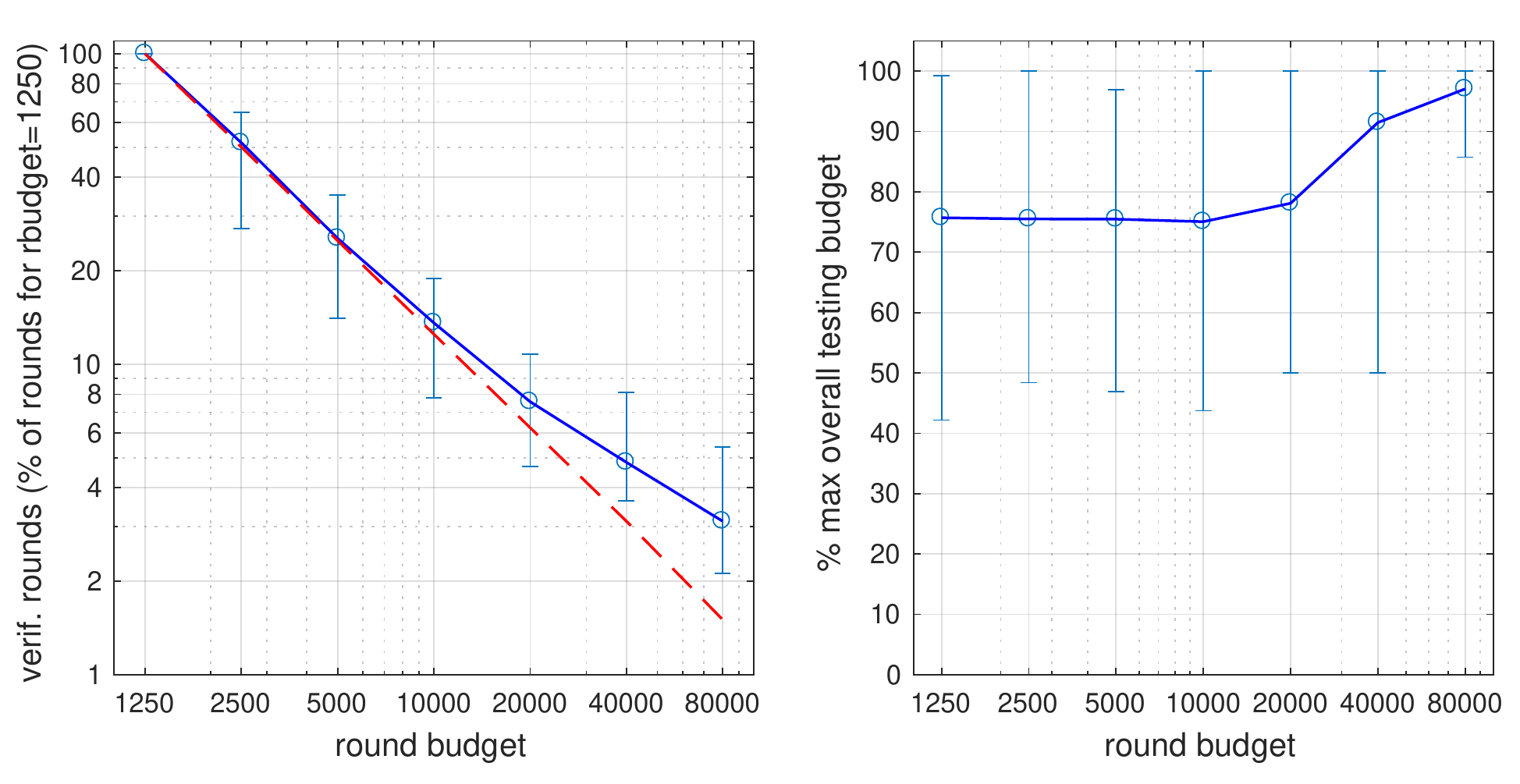}
    
    \begin{footnotesize}
    \hspace*{1cm}(a)\hspace*{5.3cm}(b)
    \end{footnotesize}
    \caption{\revised{[RQ3]} Effect of varying the \acronym\ round budget on
    (a)~the number of verification rounds; and (b)~a normalised measure of the overall testing budget (see main text for details). The plots show mean values and ranges over 10~randomly selected verification scenarios \revised{with unit component testing costs $\mathit{cost}_1=\mathit{cost}_2=\ldots=\mathit{cost}_m=1$}.
    \label{fig:round-budget-analysis}}
\end{figure*}

\subsection{Research question RQ3 \label{subsect:RQ3}}

The round testing budget $\mathit{rbudget}$ is a key parameter of \acronym. 
Large $\mathit{rbudget}$ values \revised{are undesirable because they lead to} all the component observations needed to complete the verification of the nonfunctional requirements \revised{being} acquired in a small number of verification rounds. \revised{This gives}  
\acronym\ limited opportunity to meaningfully adapt its partitioning of the round budget to the system and requirements being verified. Small $\mathit{rbudget}$ values are equally 
undesirable \revised{because they}  yield only a few additional observations in each round.  
\revised{As} such, they \revised{provide} insufficient information to properly guide the round-budget partitioning in the early verification rounds, \revised{and lead to } 
large numbers of verification rounds, \revised{which} 
can be computationally expensive because of the formal verification with confidence intervals step of \acronym.

To analyse these effects of $\mathit{rbudget}$, we randomly selected five of the TAS verification scenarios and five of the WebApp verification scenarios from Section~\ref{subsect:RQ1}, and we used \acronym\ to verify the nonfunctional requirements of the two systems for each round budget value in the set $\mathit{RB}=\{1250, 2500, 5000, 10000, 20000,$ $40000, 80000\}$. The experimental results are presented in Figure~\ref{fig:round-budget-analysis}. First, the graph from  Figure~\ref{fig:round-budget-analysis}(a) shows the \revised{number of verification rounds required when the verification was carried out using each of the round budgets from $\mathit{RB}$.}
The dashed line from this graph shows what the ``ideal'' effect of increasing the round budget would look like, i.e., a halving of the number of verification rounds each time when $\mathit{rbudget}$ is doubled, from the baseline of $100\%$ for $\mathit{rbudget}=1250$ to $50\%$ of that baseline for $\mathit{rbudget}=2500$, $25\%$ for $\mathit{rbudget}=5000$, etc. In reality, the number of verification rounds is increasingly above the ideal value as $\mathit{rbudget}$ grows, until it is above this ideal value for all 10 verification scenarios both for $\mathit{rbudget}=40000$ and for  $\mathit{rbudget}=80000$. This indicates that very large $\mathit{rbudget}$ values increase the overall testing budgets required by \acronym---a finding that is further confirmed by Figure~\ref{fig:round-budget-analysis}(b), which shows how the overall testing budget necessary to complete the verification grows with $\mathit{rbudget}$. 

\begin{figure*}
\centering
    \includegraphics[trim=5mm 87mm 8mm 95mm, clip, width=0.65\hsize]{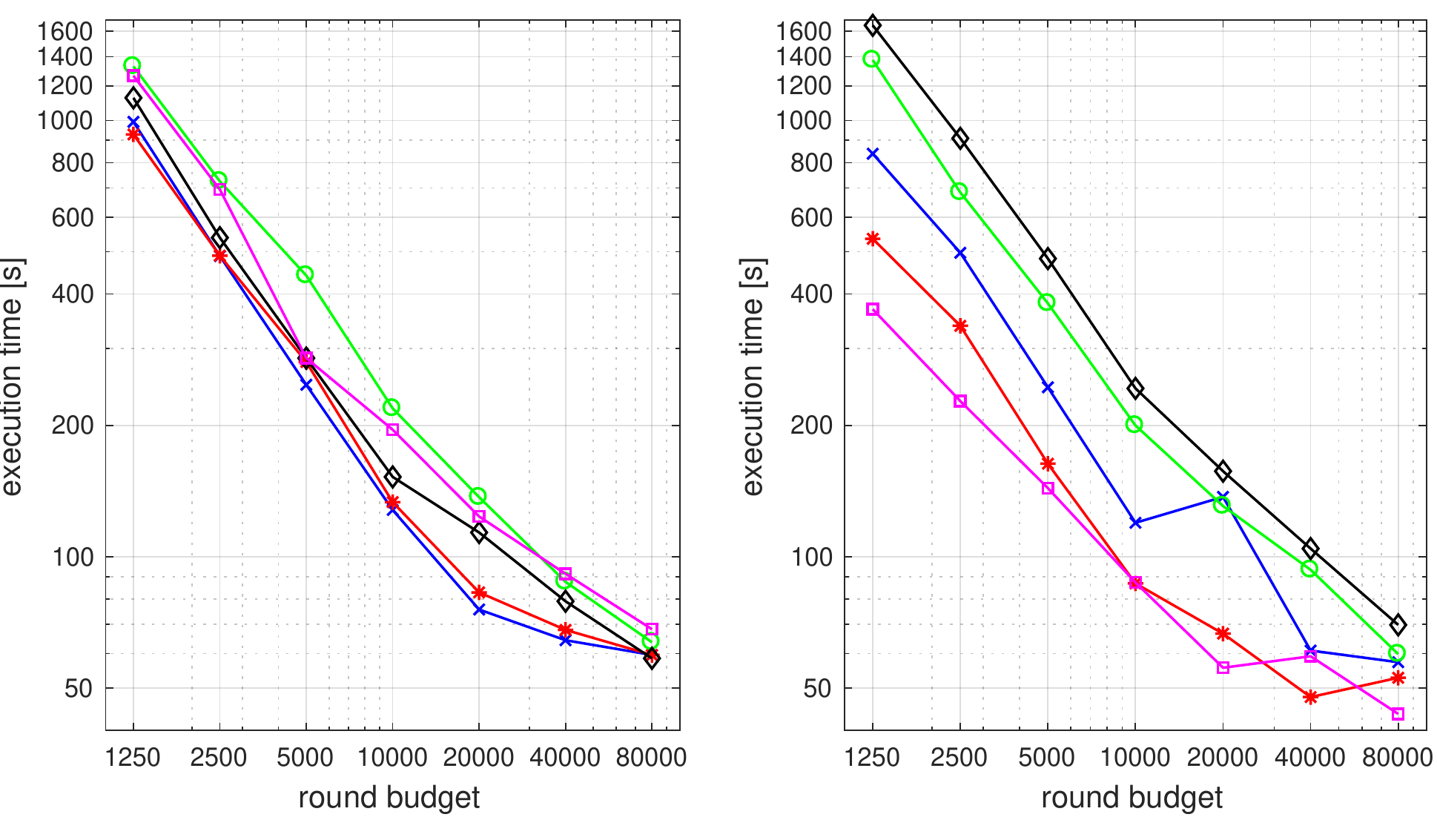}
    
    \begin{footnotesize}
    \hspace*{1.4cm}(a) TAS\hspace*{4.5cm}(b) WebApp
    \end{footnotesize}
    \caption{\revised{[RQ3]} Effect of varying the round budget on the \acronym\ execution time (experiments carried out on a c5.2xlarge Windows Server 2019 Amazon EC2 instance with 3.00GHz Intel(R) Xeon(R) Platinum 8124M CPU, and 16 GB of memory \revised{assuming unit component testing costs $\mathit{cost}_1=\mathit{cost}_2=\ldots=\mathit{cost}_m=1$}).
    \label{fig:round-budget-execution-times}}
\end{figure*}

To summarise the very different overall testing budgets required for our 10 randomly selected verification scenarios in a consistent way, Figure~\ref{fig:round-budget-analysis}(b) considers the overall testing budgets $b_1$, $b_2$, \ldots, $b_7$ associated with each verification scenario and the seven $\mathit{rbudget}$ values from $\mathit{RB}$, finds $b_\mathsf{max}=\max \{b_1,b_2,\ldots,b_7\}$, and computes the percentages of $b_\mathsf{max}$ that $b_1$, $b_2$, \ldots, $b_7$ correspond to, i.e., $\mathit{pb}_1=100b_1/b_\mathsf{max}$, $\mathit{pb}_2=100b_2/b_\mathsf{max}$, \ldots, $\mathit{pb}_7=100b_7/b_\mathsf{max}$. These ``normalised'' budgets show the round budget for which \acronym\ requires the highest overall testing budget (e.g., $\mathit{pb}_7=100$ means that the highest overall testing budget is needed when $\mathit{rbudget}=80000$), and how the testing budgets for other $\mathit{rbudget}$ values compare to that (e.g., $\mathit{pb}_1=75$ means that the overall testing budget for $\mathit{rbudget}=1250$ is 75\% of the highest overall testing budget). Figure~\ref{fig:round-budget-analysis}(b) shows how the mean of these normalised budgets increases from $\mathit{pb}_1=75.5\%$ for $\mathit{rbudget}=1250$ to $\mathit{pb}_7=97\%$ for $\mathit{rbudget}=80000$. The variability of the budget values is very large across the 10~verification scenarios from our experiments, except for the largest round budget $\mathit{rbudget}=80000$, which indicates that this round budget is consistently too large across the majority of the scenarios. 

While the effects of using very large $\mathit{rbuget}$ values are clear in Figure~\ref{fig:round-budget-analysis}, noticing the effects of small $\mathit{rbuget}$ values requires a more careful analysis of the experimental results. A first observation we can make is that the experiments with the smallest $\mathit{rbudget}$ values of $1250$, $2500$ and $5000$ used the largest number of verification rounds (as expected, see Figure~\ref{fig:round-budget-analysis}a) without delivering smaller mean overall testing budgets than the experiments for $\mathit{rbudget}=10000$ (see Figure~\ref{fig:round-budget-analysis}b). In fact, the numerical results show a very slight decrease in the mean overall testing budgets from $75.7\%$ for $\mathit{rbudget}=1250$ to $75.5\%$ for $\mathit{rbudget}=2500$, $75.47\%$ for $\mathit{rbudget}=5000$ and $75.05\%$ for $\mathit{rbudget}=10000$. Thus, \revised{Figure~}\ref{fig:round-budget-analysis}\revised{b illustrates the first undesirable effect of using} small $\mathit{rbudget}$ values, \revised{namely an} increase \revised{in} the cost of the component testing without enabling \acronym\ to adapt its partition of the round budget more effectively. 

The second undesirable effect of using small $\mathit{rbudget}$ values is visible in Figure~\ref{fig:round-budget-execution-times}, which depicts the end-to-end verification times for each of the five TAS verification scenarios and each of the five WebApp verification scenarios we used for the experiments described in this section. As shown by the logarithmic-scale graphs from this figure, the \acronym\ execution times approximately double each time the round budget is halved from $\mathit{rbudget}=10000$ to $\mathit{rbudget}=5000$, to $\mathit{rbudget}=2500$ and, finally, to $\mathit{rbudget}=1250$. Even when \acronym\ is used at design time and execution times of close to 30~minutes (for $\mathit{rbudget}=1250$) are acceptable, the results from Figure~\ref{fig:round-budget-analysis}b show that such long execution times yield no benefit, so very small $\mathit{rbudget}$ are not recommended.
\revised{As a side comment, we note that the VERACITY execution time does not always decrease monotonically in Figure~}\ref{fig:round-budget-execution-times}\revised{b. This is due to variations in the computation time required to calculate the confidence intervals in step~1 of the VERACITY verification process (cf.~Figure~}\ref{fig:approach}\revised{), to the stochasticity of the observations from the component testing performed in step~3 of this process, etc.}

\subsection{Threats to validity \label{subsect:threats}}

\emph{Construct validity} threats may arise due to assumptions made when modelling the systems from our case studies. To mitigate these threats, we used models and nonfunctional requirements based on established case studies from the research
literature~\cite{calinescu2011dynamic,baresi2007validation,filieri2012formal}. 

\emph{Internal validity} threats may be caused by bias in establishing cause-effect relationships in our experiments. To limit these threats, we assessed \acronym\ for large numbers of verification scenarios with randomly generated nonfunctional requirements bounds for each research question and each case study: 378 verification scenarios for research question RQ1, 378 verification scenarios for RQ2, and 70 verification scenarios for RQ3. Furthermore, as explained at the beginning of Section~\ref{subsect:RQ1}, we ensured that these experiments included scenarios where the requirements were satisfied, where they were violated, and where some requirements were satisfied and others were violated---both by a wide margin and by a narrow margin. Finally, we enable replication by making all experimental results available on our project's website.

\emph{External validity} threats may exist if the verification of the nonfunctional requirements of other software systems cannot be expressed in the format from our problem definition in Section~\ref{subsect:problem}. We limited these threats by ensuring that \acronym\ supports the verification of systems whose behaviour is modelled using parametric Markov chains encoded in the widely used modelling language of the PRISM model checker~\cite{prism}, with nonfunctional requirements specified in the established temporal logic PCTL~\cite{Andova2004,Ciesinski2004,Hansson1994}. Parametric Markov models are increasingly used to model software systems including service-based systems~\cite{gerasimou2018synthesis}, software product lines~\cite{ghezzi2013model}, software controllers of cyber-physical systems~\cite{calinescu2017engineering}, and multi-tier software architectures~\cite{calinescu2019efficient}. Nevertheless, additional experiments  are needed to establish the applicability and feasibility of \acronym\ in domains with characteristics different from those used in our evaluation.

\revised{Another external validity threat may arise if the verification of other software systems requires the use of larger Markov models than those that we used to evaluate VERACITY. To mitigate this threat, we used systems and models proposed by other projects in established software engineering venues~}\cite{baresi2007validation,filieri2012formal}\revised{. Additionally, we assessed the level of this threat by comparing the size of these models to that of the discrete-time Markov chains (whether parametric or not) from all the research papers:}
\begin{itemize}
    \item \revised{published within the past five full years (2016--2020) in the \emph{Journal of Systems and Software} and in all the CORE2020\footnote{\url{https://www.core.edu.au/conference-portal}} rank A* software engineering journals (i.e., \emph{IEEE Transactions on Software Engineering} and \emph{ACM Transactions on Software Engineering and Methodology}) and conferences (i.e., the \emph{International Conference on Software Engineering} and the \emph{European Software Engineering Conference and the ACM SIGSOFT Symposium on the Foundations of Software Engineering});}
    \item \revised{that mention the size of the DMTCs used for evaluation, or provide these DTMCs or access to them on a project website or in a project repository.}
\end{itemize}
\revised{Across the eight papers that met these criteria  ~}\cite{calinescu2017engineering,su2016reliability, franco2016improving,llerena2018verifying, wang2018towards, nakagawa2019expression, afzal2019performance,7355393}, models \rerevised{not larger than} our TAS model (i.e., 10 states) are used for evaluation in five papers, and models smaller than or of the same size as our shopping application model (i.e., 17 states) are used for evaluation in all eight~papers. While five papers also use models larger than ours for evaluation, these are not models of software systems with multiple components (to which VERACITY is applicable) but models of algorithms, programs and logical gates, or synthetic models that do not correspond to an actual software system. This analysis (whose detailed results are provided on our project's website) shows that abstraction allows many software engineering projects to usefully exploit models of similar size to the models used for evaluation in our paper. Nevertheless, larger mode\rerevised{l}s are sometimes required, and \rerevised{therefore we carried out an additional case study in which we applied VERACITY to an 18-parameter, 41-state discrete-time Markov chain modelling a larger component-based software system taken from }\cite{calinescu2019efficient}\rerevised{. This case study, reported on our project website }\url{https://www.cs.york.ac.uk/tasp/VERACITY}\rerevised{, indicates that VERACITY can also handle models with larger numbers of parameters and states than those presented earlier in this section}.

\section{Related work \label{sect:related}}

Within the past decade, the study of uncertainty in the modelling, analysis and verification of complex systems has attracted significant attention from the research community. As such, the existence of different classes of uncertainty is now widely recognised \cite{RamirezSEAMS12,Giese2014,ICPE14,Walker}, and the research literature provides multiple definitions of uncertainty. Most of these definitions classify uncertainties depending on their: (i)~\emph{level} (ranging from determinism to complete ignorance); (ii)~nature (aleatory or epistemic);  and (iii)~source (in the structure or parameters of models, associated with changes in the operational environment or with the dynamics in the availability of resources, or due to changes in the user goals) \cite{Walker, Garlan2010,Esfahani13,ICPE14}.

When dealing with the verification of performance, reliability, or other nonfunctional requirements, the term uncertainty is often defined in terms of aleatory and epistemic uncertainty.  The aleatory variability of parameters and indices is typically captured using  stochastic modelling notations, while the epistemic uncertainty, which refers to the behavior of system portions that are intrinsically unknown, requires ad-hoc methods. 

The common goal of these methods is to introduce analysis methodologies able to produce satisfactory results even in presence of such types of uncertainty. For example, \cite{CatiaQoSA13,Grunske-QoSA11,GrunskeJSS12} propose methods that can be used at design time, to identify  software architectures, configurations or component compositions that satisfy a given set of nonfunctional requirements. These approaches take into account the epistemic uncertainty associated with the parameters of models, and use probability distributions to model this uncertainty. To this end, they obtain samples of the uncertain parameter values, and evaluate the robustness of a software system under uncertainty 
by running Monte-Carlo simulations that use these empirical probability distributions.

More recently, \cite{das20} focuses on understanding the influence of configuration options on performance
and proposes an  approach based on probabilistic programming that explicitly models uncertainty for option influences and provides both a scalar and a confidence interval
for each prediction of a configuration’s performance. A method that uses mathematical formulas for incorporating and evaluating epistemic uncertainty of the input parameters of queueing models is presented in~\cite{ANTONELLI2020746}. A similar approach to the study of uncertainty propagation in reliability models is introduced in~\cite{Mishra2013}.

A different philosophy in dealing with uncertainty involves the adoption of self-adaptation in software systems. The number of studies that consider uncertainty in self-adaptive systems has increased in recent years. A typical example is the use of  probabilistic run-time models, such as Markov decision processes~\cite{calinescu2017synthesis,Moreno:2015} 
and parametric stochastic models~\cite{calinescu2018efficient} 
to reason about uncertainty and change when making adaptation decisions. 
The approaches introduced in \cite{Weyns2015,musil2017patterns} employ self-adaptation to cope with uncertainty. The approach from \cite{Weyns2015} proposes a combination of adaptation and evolution of the software to make its behavior resilient to uncertainty, which in turn entails that the software system is \emph{sustainable}, while \cite{musil2017patterns} focuses on the uncertainty surrounding the execution of cyber-physical production systems. 
A different approach can be found in~\cite{Simca19}, where a control-theoretic approach is adopted to handle uncertainty in self-adaptive software systems. 
Furthermore, the need for software systems to operate well under the existing uncertainties is among the main waves that have advanced the research on self-adaptive systems \cite{Weyns2017}, although a \emph{perpetual assurance} of goal satisfaction in self-adaptive systems is still an open research challenge \cite{perpetualassurances}. Most of these results consider uncertainty in the decision-making process and propose adaptation approaches that guarantee the quality requirements under different and (possibly) unknown types of changes.

Our work lies in the area of the reduction of parametric epistemic uncertainty and introduces an adaptive uncertainty reduction heuristic for performance and dependability software engineering. The proposed heuristic is integrated into a new iterative approach that exploits the adoption of formal verification with confidence intervals.

One of the key aspects of the proposed approach consists of the identification of the system component for which additional data---to be obtained through testing---are needed. The selection of components to be tested in each iteration is based on a combination of factors that include the sensitivity of the model to variations
in the parameters of different components, and the overheads  of unit-testing each of these components.
Reducing the cost of the (reliability) testing phase by selecting key components to test is a topic that has been analysed in the literature. For example, \cite{Garg2011} tackles the question ``When to stop testing''  by focusing on reliability and discussing the challenges and the potentials related to existing software reliability models.
Classical approaches in this domain are based on operational profile~\cite{Musa}, however operational profile is often unknown and subject to changes. To overcome this problem, \cite{Russo16}~proposes an adapting testing schema that iteratively learns from test execution results as they become available, and, based on them, allocates test cases to the most sensible parts. The assessment is then performed adopting a second sampling strategy that provides the interval estimate of the reliability computed during testing. A different approach that focuses on the allocation of testing resources under uncertain conditions is presented in~\cite{Russo18}. In this approach, a multi-objective debug-aware and robust optimization problem under uncertainty of data is used to evaluate of alternative trade-offs among reliability, cost, and release time. 

Compared to these approaches, \acronym\ has the major advantage that it uses formal quantitative verification to compute confidence intervals for the relevant nonfunctional properties of the system under verification. As such, our approach mitigates the risk of generating inaccurate single-point estimates for these properties. Furthermore, its use of Markov models makes \acronym\ applicable to multiple classes of systems for which such models are extensively used, as detailed in Section~\ref{subsect:threats}.

\section{Conclusion \label{sect:conclusion}}

We presented \acronym, a tool-supported approach for the efficient verification of nonfunctional requirements under uncertainty. \acronym\ operates by acquiring information about the components of the verified system through testing them individually over a number of verification rounds. A user-defined testing budget specifies the amount of testing performed in each round, and the partition of this budget among system components is adapted from one round to the next in order \revised{to }complete the verification process with a low overall testing cost. The heuristic used to compute this adaptive partition considers factors such as the sensitivity of the verified requirements to the parameters associated with different components, and the different cost (e.g., time, price or risk) of testing these components. The evaluation of \acronym\ in case studies from the areas of service-based systems and web applications showed that, on average, it significantly reduces the overall testing cost required to complete the verification process compared to uniformly partitioning the testing budget across all system components.

In future work, we plan to expand the set of factors underpinning our \acronym\ round-budget partitioning, in order to further improve its efficiency. One such additional factor that we are considering is the level of epistemic uncertainty associated with each component: in each round, a larger fraction of the testing budget should be allocated to components with higher levels of epistemic uncertainty, i.e., to those for which fewer observations are already available. We envisage that augmenting our heuristic with this factor will extend the applicability of \acronym\ to verification scenarios in which observations about a subset of the system components are already available at the beginning of the verification process (e.g., from previous testing of those components), and we plan to carry out additional case studies to validate this hypothesis. 

Another important direction of future research for our project is to consider the scenario\revised{s} in which \revised{(i)~the imbalance between the component testing costs is much higher than the 1:5 ratio considered in our experiments so far; and/or (ii)}~some of the parameters that the verified requirements depend on are associated with the operational profile of the system, i.e., with parameters whose epistemic uncertainty cannot be lowered by testing the components of the system. \revised{For the second scenario,} examples of such parameters include the number of requests received by a web server in one hour, and the probabilities of these requests being of different types. To some extent, \acronym\ could handle this scenario by associating such parameters with an ``operational profile component'' that is assigned an infinite testing cost. Because this ``component'' will never be tested, the verification problem may be undecidable, in which case \acronym\ will (correctly) terminate with a `budget exhausted' outcome. However, this outcome will only be produced after significant testing effort, some of which could be avoidable by noticing---before using all the testing budget---that the operational profile uncertainty renders the verification problem undecidable. We plan to extend \acronym\ with the ability to report an `undecidable' outcome (without exhausting the testing budget) in this important verification scenario.

\revised{Last but not least, the verification problem tackled by VERACITY can be generalised in multiple ways. For example, cost can be considered a multi-dimensional entity with separate elements for time, monetary cost, etc.; in \rerevised{such a case,} the budget would also be a tuple with these elements. 
As another example, \rerevised{it} may be of interest to use different confidence levels $\alpha_1$, $\alpha_2$, \ldots, $\alpha_n$ for the $n$ requirements from~}\eqref{eq:reqs}\revised{. To handle this variant of the problem, we envisage that VERACITY will need to be augmented with the ability to acquire more observations for the system components with parameters that influence the requirements associated with high confidence levels than for the other components. We plan to explore this hypothesis in our future work.}

\section*{Appendix~A}

\paragraph{Markov chain construction from UML activity diagrams} A parametric discrete-time Markov chain~\eqref{eq:mc} can be derived from the UML activity diagram of a software system by following the step-by-step process summarised below:
\begin{enumerate}
    \item Construct the state set $S$ consisting of a state for each activity node from the UML activity diagram, plus an initial state $s_0$ and an ``end'' state $s_\mathsf{end}$ associated with the initial and final nodes of the activity diagram, respectively. For each state $s\in S$, let $\mathit{node}(s)$ represent the activity diagram node corresponding to state $s$.
    \item Set $\mathbf{P}(s,s')=1.0$ for every pair of states $s,s'\in S$ for which the node reached immediately after $\mathit{node}(s)$ in the activity diagram (i.e., without traversing activity diagram nodes associated with states from $S\setminus\{s,s'\}$) is \emph{always} $\mathit{node}(s')$.
    \item Set $\mathbf{P}(s_\mathsf{end},s_\mathsf{end})=1.0$.
    \item Associate an unknown transition probability $\mathbf{P}(s,s')$ with each pair of states $s,s'\in S$ for which $\mathit{node}(s')$ can be reached from $\mathit{node}(s)$ by traversing only decision nodes from the activity diagram.
    \item Set $\mathbf{P}(s,s')=0$ for every other pair of states $s,s'\!\in\! S$.
    \item Assemble the atomic proposition set \[\mathit{AP}\!=\!\{\mathit{start}, \mathit{end}\}\cup\{\mathit{name}(\mathit{node}(s))\!\mid\! s\!\in\! S\setminus\{s_0,s_\mathsf{end}\}\}\]
    where $\mathit{name}(\mathit{node}(s))$ is a unique name for the activity node $node(s)$.
    \item Set $L(node(s))=\mathit{name}(\mathit{node}(s))$ for every state $s\in S\setminus \{s_0,s_\mathsf{end}\}$, $L(s_0)=\mathit{start}$, and $L(s_\mathsf{end})=\mathit{end}$.
\end{enumerate}
Further details about this process are available in~\cite{calinescu2013using,ghezzi2013managing}.

\section*{Acknowledgements}
This project has received funding from the Assuring Autonomy International Programme, the UKRI project EP/V026747/1 `Trustworthy Autonomous Systems Node in Resilience', and the ORCA-Hub PRF project `COVE'.

\end{document}